\documentclass[a4paper,11pt]{article}
\usepackage{graphicx,rotating,wrapfig,float}
\usepackage[T1]{fontenc}
\usepackage{units}
\usepackage{cleveref}
\usepackage[margin=10pt,font=small,format=plain,labelfont=bf]{caption}
\usepackage{enumerate}
\usepackage{amsmath}
\usepackage{authblk}

\begin{document}



%

\title{\bf Status of the ArDM Experiment: First results from gaseous argon operation in deep underground environment}

\author[1]{A.~Badertscher}
\author[1]{F.~Bay}
\author[3]{N.~Bourgeois}
\author[1]{C.~Cantini}
\author[2]{M.~Daniel}
\author[1]{U.~Degunda}
\author[1]{S.~Di~Luise}
\author[1]{L.~Epprecht}
\author[1]{A.~Gendotti}
\author[1]{S.~Horikawa}
\author[1]{L.~Knecht}
\author[1]{D.~Lussi}
\author[3]{G.~Maire}
\author[2]{B.~Montes}
\author[1]{S.~Murphy}
\author[1]{G.~Natterer}
\author[1]{K.~Nikolics}
\author[1]{K.~Nguyen}
\author[1]{L.~Periale}
\author[3]{S.~Ravat}
\author[1]{F.~Resnati}
\author[2]{L.~Romero}
\author[1]{A.~Rubbia\thanks{andre.rubbia@cern.ch}}
\author[2]{R.~Santorelli}
\author[1]{F.~Sergiampietri}
\author[1]{D.~Sgalaberna}
\author[1]{T.~Viant}
\author[1]{S.~Wu}

\affil[1]{ETH Zurich, Institute for Particle Physics, Schafmattstrasse 20, CH-8093 Z\"{u}rich, Switzerland}
\affil[2]{CIEMAT, Div. de F{\'\i}sica de Particulas, Avda. Complutense, 22, E-28040, Madrid, Spain}
\affil[3]{CERN, PH/DT group, 1211 Geneve 23, Switzerland}

\renewcommand\Authands{ and }
\renewcommand\Authfont{\scshape\small}
\renewcommand\Affilfont{\itshape\tiny}
\date{\vspace{-5ex}}

\maketitle
\begin{center}
The ArDM Collaboration
\end{center}

\begin{abstract}
The Argon Dark Matter (ArDM) 
experiment is a ton-scale liquid argon (LAr) double-phase time projection chamber designed for direct Dark Matter searches. Such a device allows to explore the low energy frontier in LAr. 
After successful operation on surface at CERN, the detector has been deployed underground and
is presently commissioned at the Canfranc Underground Laboratory (LSC). In this paper, we describe the status of the installation and present first results on data collected in gas phase.
\end{abstract}

\section{Introduction}
Astronomical observations give strong evidence for the existence of non-luminous and non-baryonic matter, presumably composed of a new type of elementary particles. The leading candidate is a cold thermal relic gas of Weakly Interacting Massive Particles (WIMPs) ~\cite{Steigman:1984ac}, which feel the gravitational interaction, but are otherwise interacting more weakly than standard weak interactions. Direct detection could be achieved by observing the energy deposited when WIMPs elastically scatter from an ordinary target material nucleus, requiring the measurement of recoils of target nuclei with kinetic energy in the range of 5--100 keV. 
Liquid argon represents a promising target with a favourable form factor and it is only sensitive to spin-independent interactions. The Argon Dark Matter 
(ArDM) experiment ~\cite{Rubbia:2005ge,Marchionni:2010fi} is using a ton-scale liquid argon (LAr) target for direct WIMP searches, operated as a double-phase time projection chamber with imaging and calorimetric capabilities.

\section{The ArDM detector}
\label{sec:2}
\subsection{Overview}
\label{sec:2.1}

The Argon Dark Matter Experiment (ArDM, CERN RE18, LSC EXP-08) is a liquid argon calorimeter/TPC with a vapour layer on top of the liquid for a double phase charge readout. It was constructed and operated at CERN \cite{Marchionni:2010fi,Epprecht:2012Diss}. In early 2012 the transfer to its final place in the underground laboratory of Canfranc (LSC) was initiated with the installation of 
the vessel and the majority of the cryogenic parts. 
The actual detector was further improved and updated to make it suitable for the operation under ground.

In February 2013 a totally rebuilt detector 
consisting of several improvements was sent and installed at LSC. It has different sub units which will be described in the following sections. Each of them was delivered as a 
pre-assembled part to Canfranc in boxes, sealed against external influences like dust and light. The final assembly then was done in the clean room of LSC. 

\begin{figure}[t]
\begin{center}
\includegraphics[width=140mm]{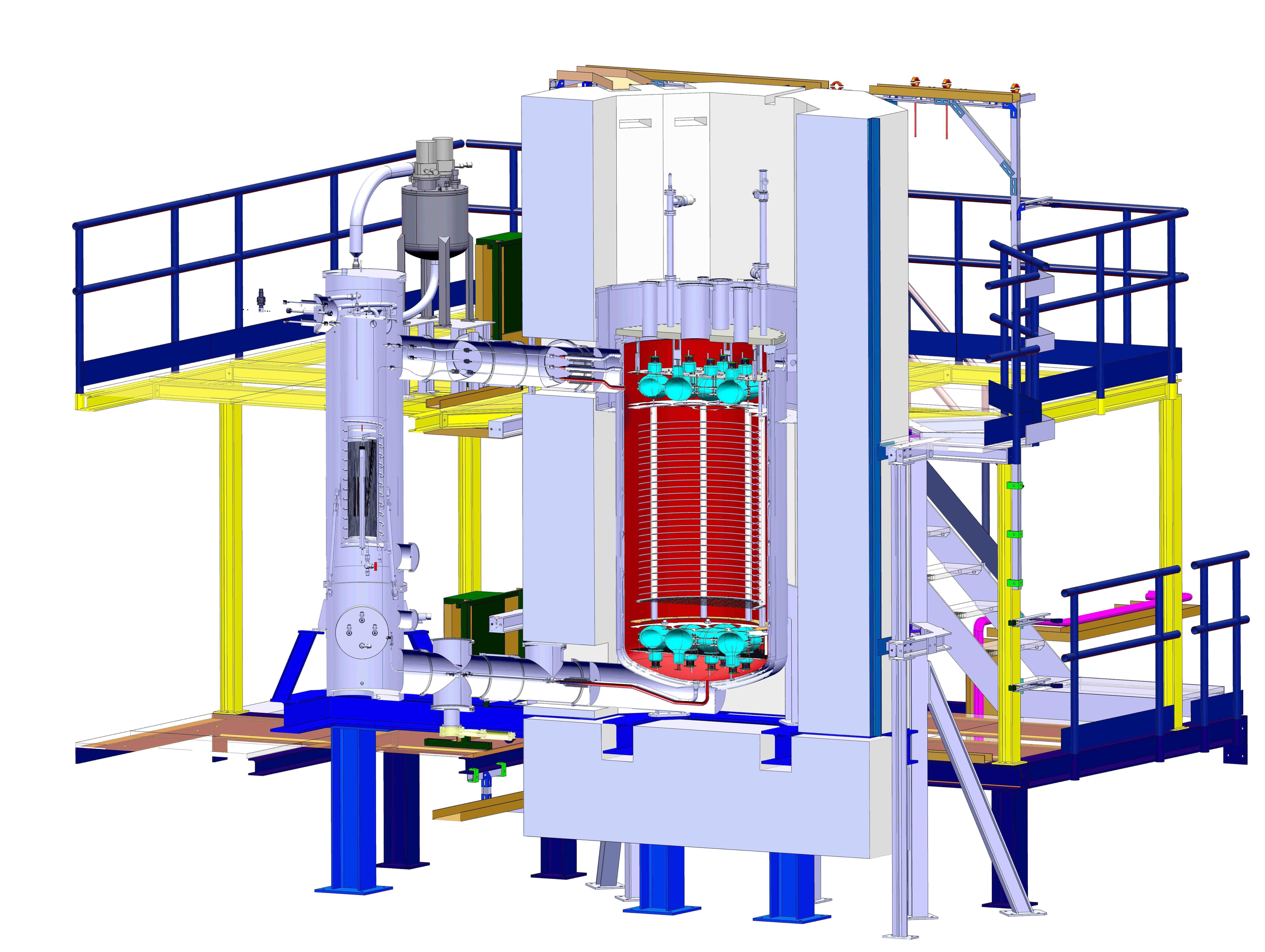}
\caption{Design of the ArDM detector: Cut view through the main vessel.}
\label{f::CAD_ArDM}
\end{center}
\end{figure}
\Cref{f::CAD_ArDM} shows a cut through the detector and its installations. For safety reasons, the actual machine is standing on a platform, one meter above ground. While on the left of \Cref{f::CAD_ArDM} the cryogenic services are visible, the actual detector on the right is surrounded by a neutron shield of polyethylene. 

It consists of two arrays with twelve 8'' photomultiplier tubes (PMTs) each: one below the fiducial volume immersed in LAr and one in the vapour phase above the surface of the liquid. The total drift length is \unit[1100]{mm} and the diameter is \unit[780]{mm}. In order to have the best possible light yield, all PMTs have been re-coated with TPB (1,1,4,4,-tetraphenyl-1,3,-butadiene) before installation. Also all the Tetratex$\unit{^{\mbox{\tiny{\textregistered}}}}$ (TTX) side reflectors, surrounding the drift volume have been re-done as described in \Cref{sec:2.1.2}. Apart of the PMTs and the side reflectors, also the top and the bottom reflectors filling the gap between the PMTs, made of PTFE sheets, newly are coated.

\subsubsection{The PMT arrays}
\label{sec:2.1.1}

\begin{figure}[t]
\begin{center}
\includegraphics[height=53mm]{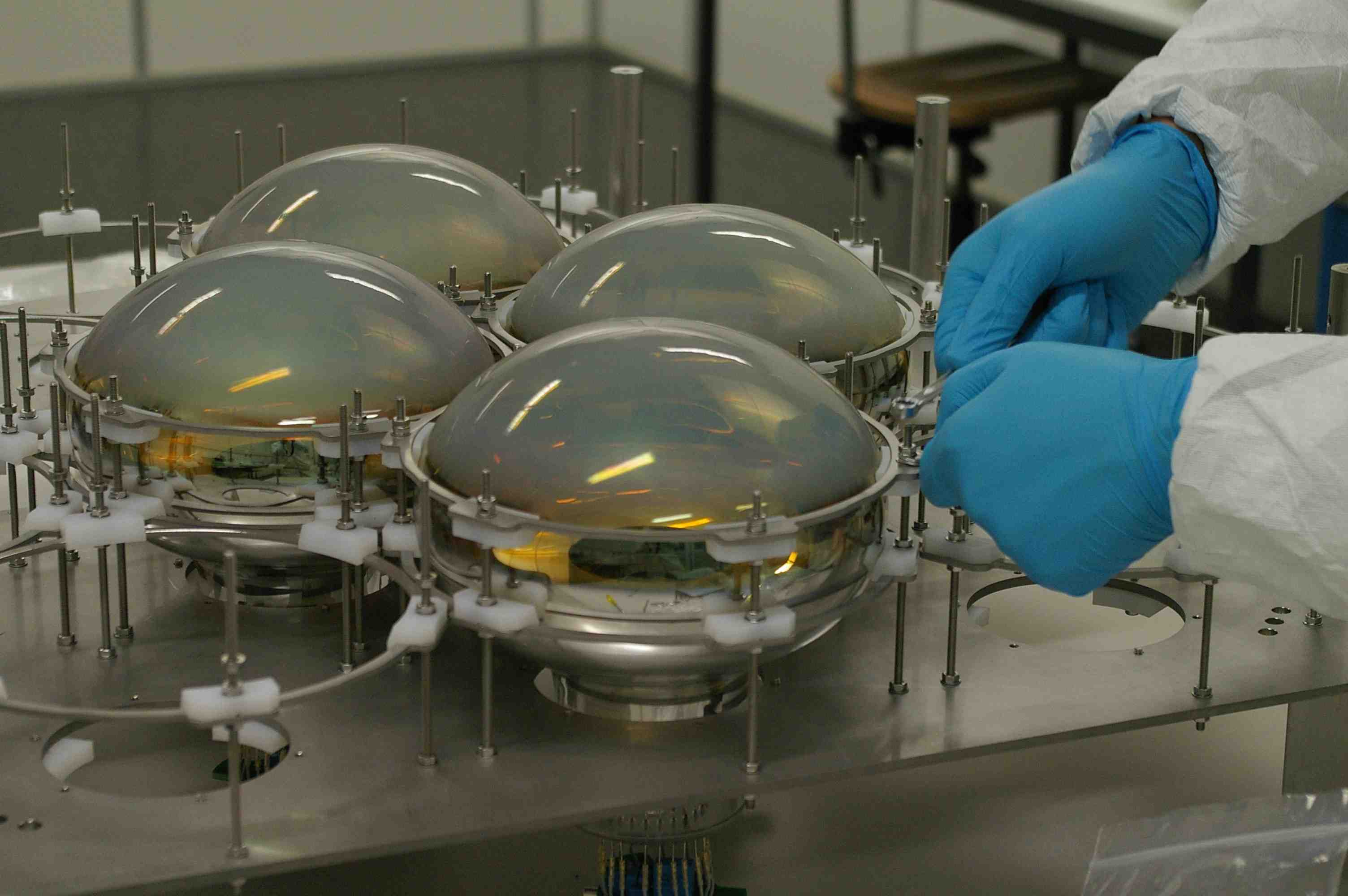}\hspace{5mm}
\includegraphics[trim=18cm 0cm 20cm 16cm, clip=true,height=53mm]{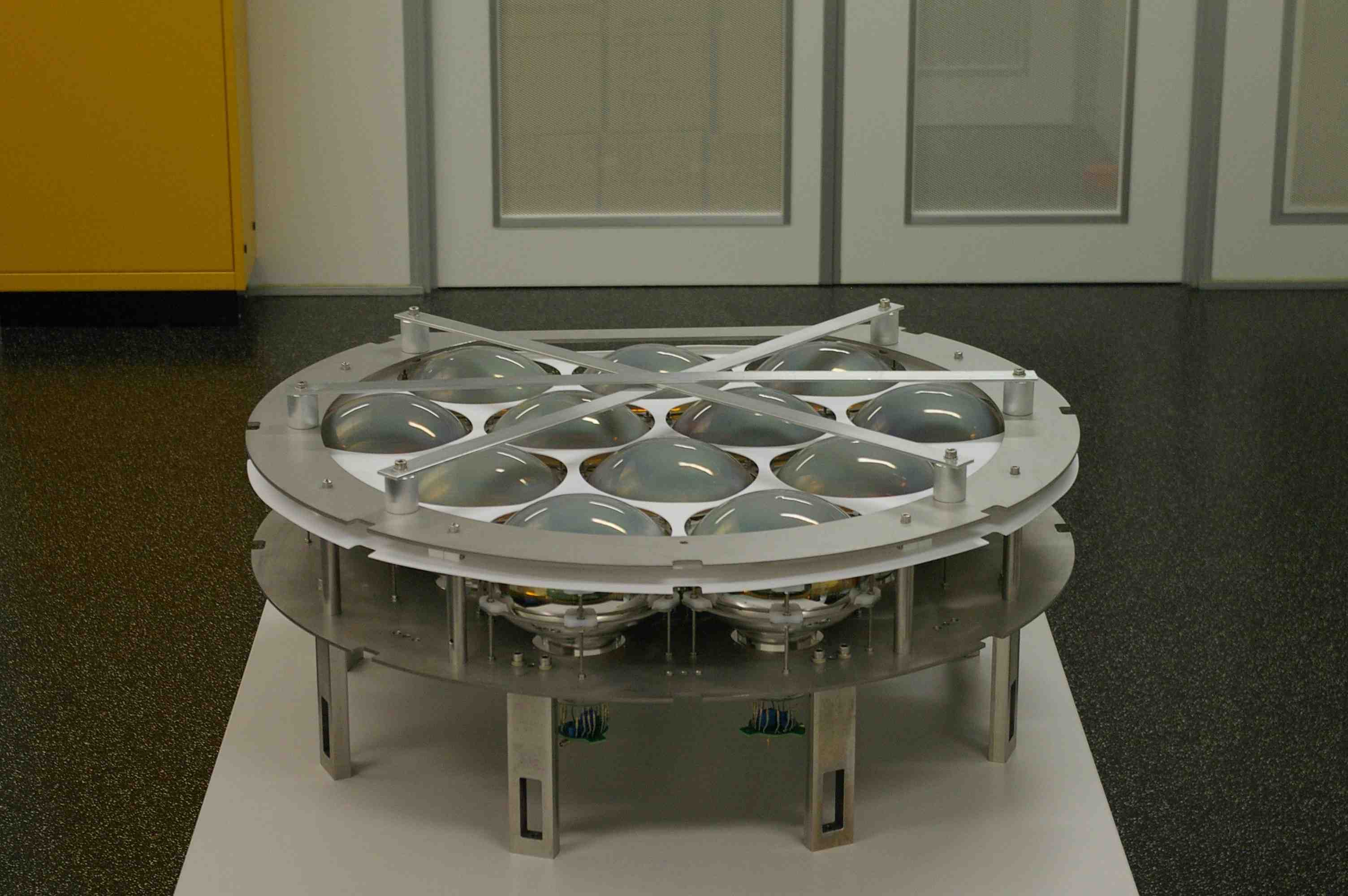}
\caption{The installation and final PMT array in the ETH clean room. The aluminium crosses over the final array are only for protection during the transport.}
\label{f_PMT_Assembly}
\end{center}
\end{figure}

The two PMT arrays consist of 12 PMTs each and are basically mirrored. 
The 8'' PMTs, Hamamatsu 5912-02MOD-LRI, coated with TPB by evaporation, are arranged in the most dense way in a hexagonal pattern. This gives a total of \unit[68]{\%} of active surface on the top and on the bottom.
These arrays have been recently constructed 
with a completely new design of the supporting structure in order to minimise the mechanical stress at cryogenic operation. 
Each PMT is individually fixed to a massive steel plate. This gives more degrees of freedom for adjustments to the thermal contractions when cooled to the temperature of LAr. Also it is relatively easy to replace a single PMT in case of a problem. 
\Cref{f_PMT_Assembly} shows the assembly of the top PMT array in a clean room. 

\subsubsection{Side reflectors}
\label{sec:2.1.2}

Since only the top and the bottom of the fiducial volume are instrumented, the light emitted to the side 
of the cylindrical fiducial volume
has to be reflected (possibly several times) 
until it reaches a photocathode. 
As the \unit[128]{nm} VUV light of argon scintillation is not 
reflected efficiently by any suitable material, 
the side reflectors are coated with TPB wavelength shifter.  

\begin{figure}[t]
\begin{center}
\includegraphics[height=70mm]{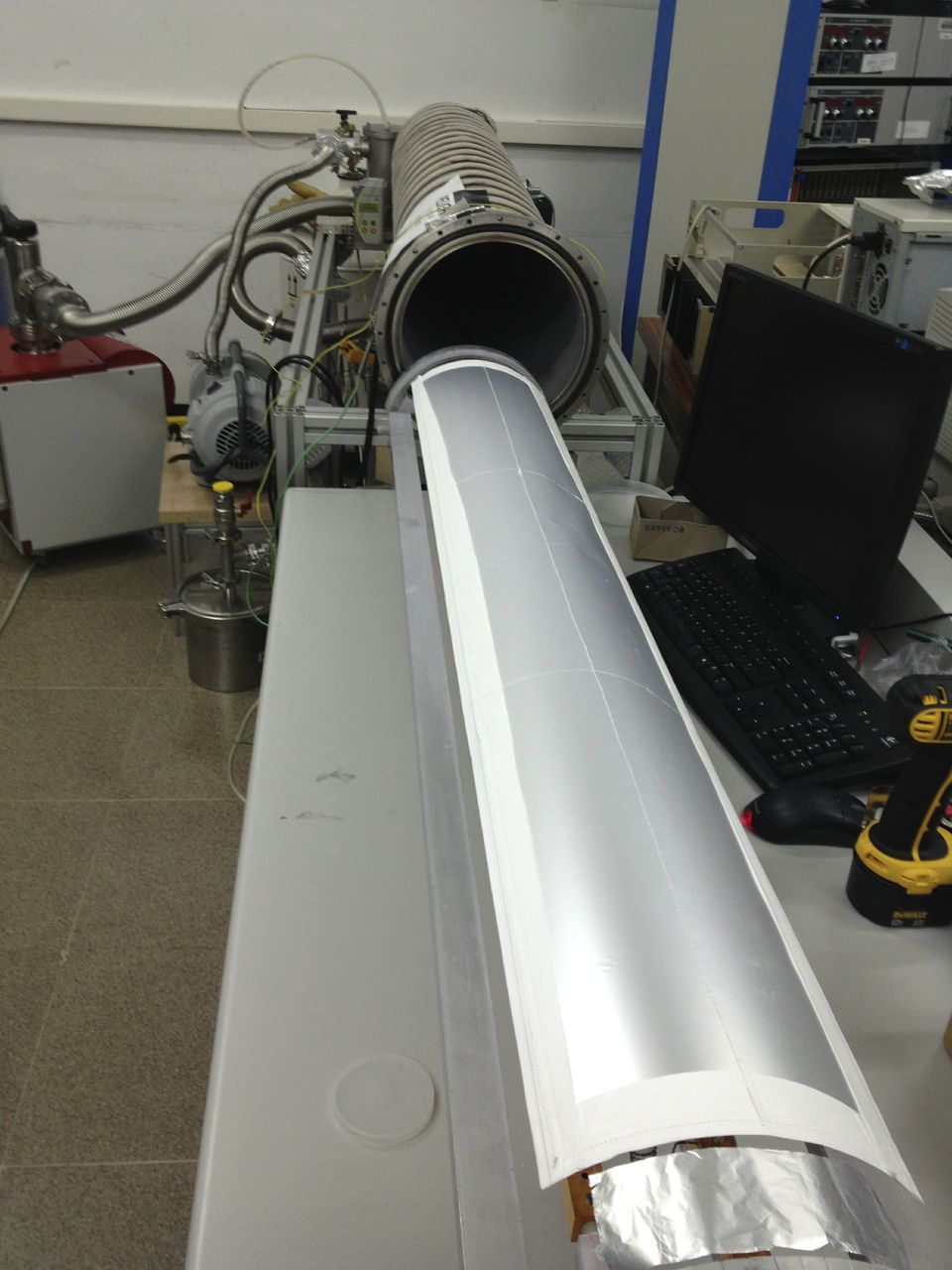}\hspace{5mm}
\includegraphics[height=70mm]{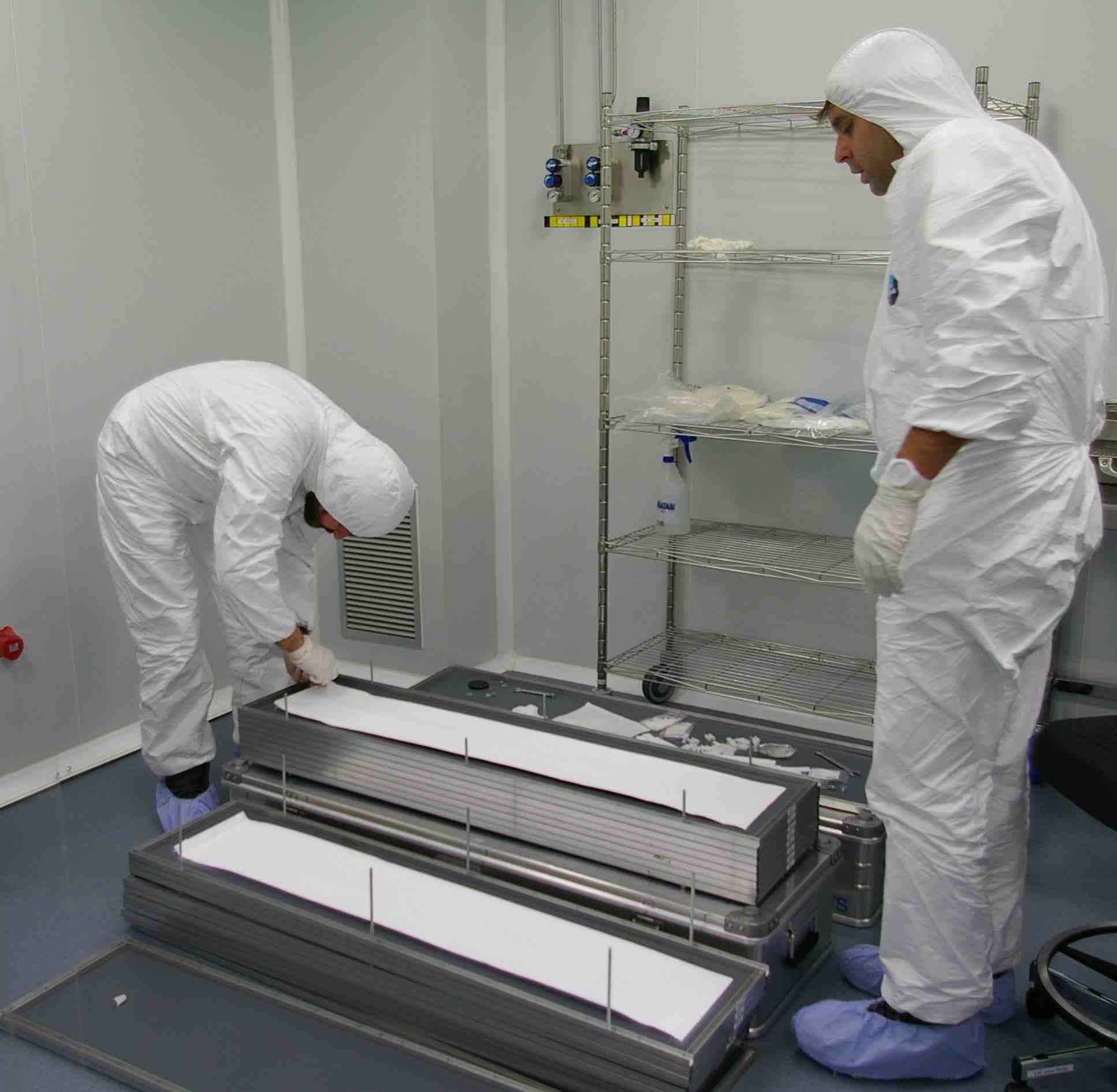}
\caption{The side reflectors were coated with TPB in the vacuum evaporator of IFIC (left) and then transported to the LSC clean room (right).}
\label{f_Reflector}
\end{center}
\end{figure}

Using the original ArDM design \cite{Boccone:2009kk}, 
the reflectors  are sheets of TTX 
with the size of $\unit[200 \times 1080 ]{mm^{2}}$. Tetratex is a fabric and favoured to a standard 
PTFE sheet because the TPB molecules attach better due to the non smooth surface. 
In order to reinforce the reflectance of the 10-mil-thick TTX foil 
a multi-layer plastic film reflector, Vikuiti{\texttrademark} ESR (Enhanced Specular Reflector) 
from 3M$\unit{^{\mbox{\tiny{\textregistered}}}}$,  is sewed on the back side of the TTX. 

The coating was done by vacuum evaporation using the evaporator developed for ArDM at CERN, and now located at IFIC in Valencia. 
The coated reflector sheets were transported to Canfranc, 
packed in aluminium boxes to protect 
them from dust and light. The unpacking in the LSC clean room can be seen in the right picture of \Cref{f_Reflector}.

\subsubsection{Final assembly and installation at LSC}
\label{sec:2.1.3}

\begin{figure}[htbp]
\begin{center}
\includegraphics[height=50mm]{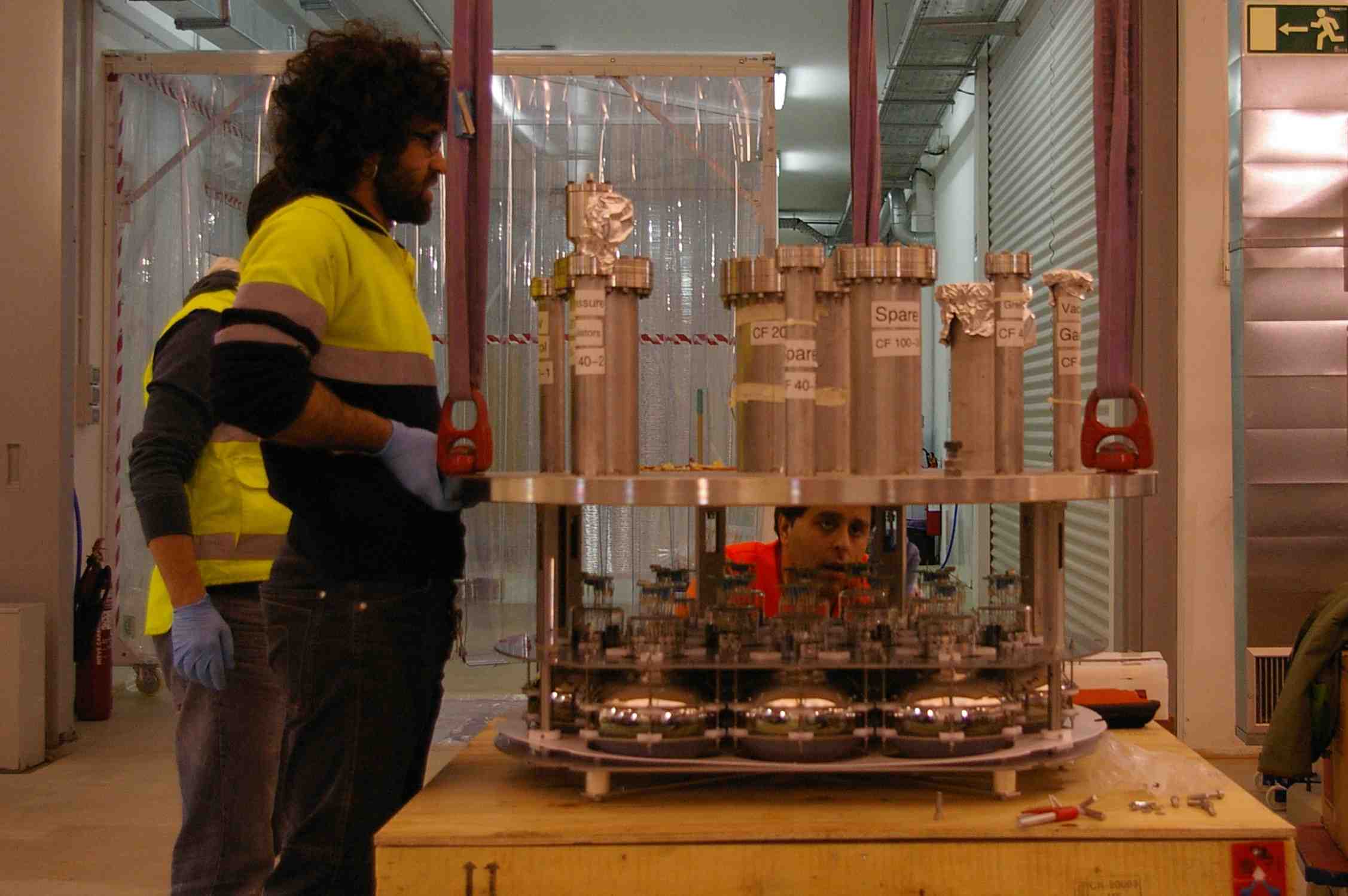}\hspace{1mm}
\includegraphics[height=50mm]{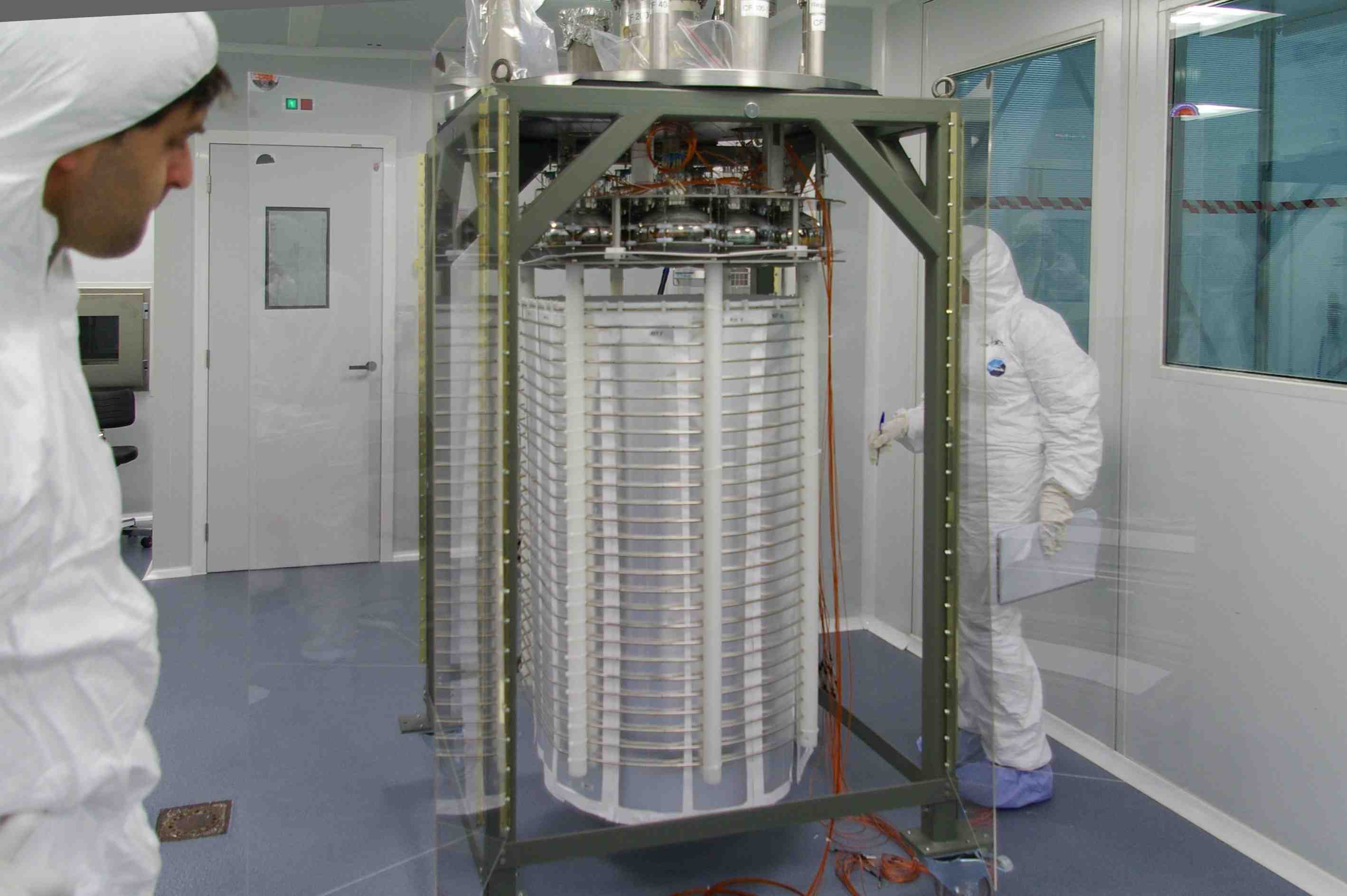}\vspace{2mm} \\
\includegraphics[height=50mm]{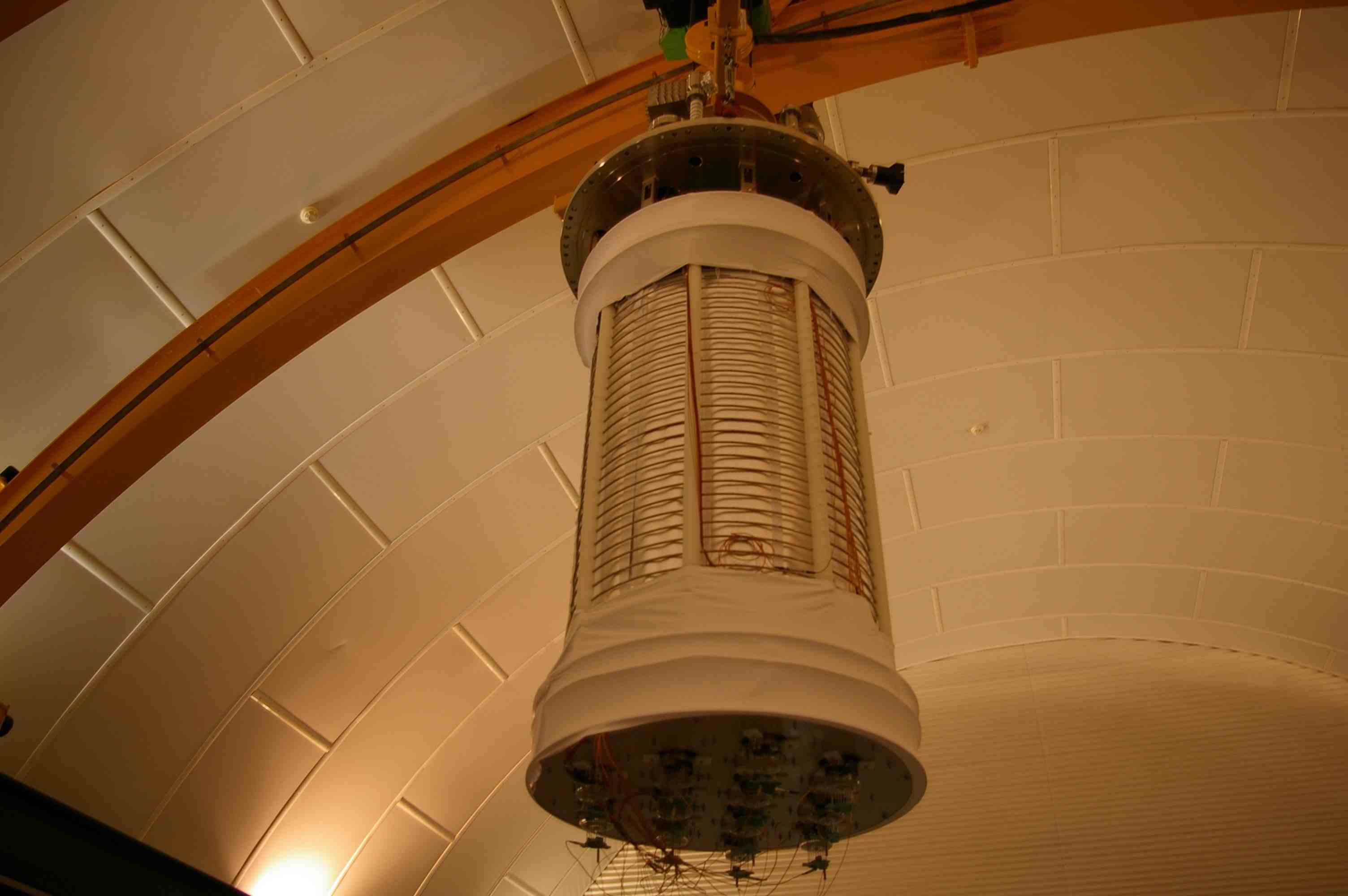}\hspace{1mm}
\includegraphics[height=50mm]{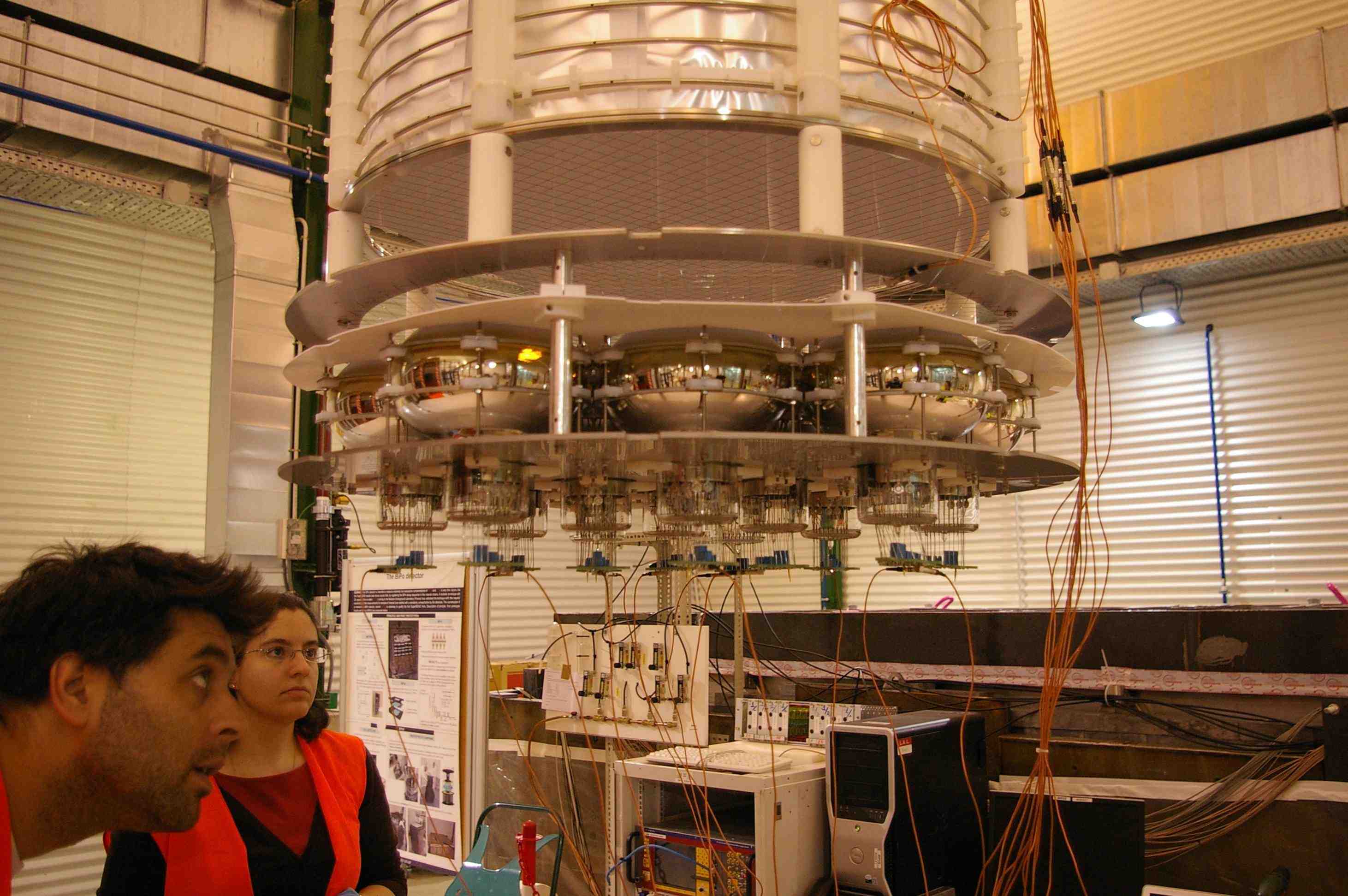}
\caption{From top left, clock wise: Fixing of upper PMT array to top flange; Installation of the drift cage and all cabling in the clean area; Installation of lower PMT array, detector back hanging from the crane; Final detector on its way back in the dewar. }
\label{f_InstallationCanfranc4}
\end{center}
\end{figure}

Finally the detector was assembled underground at LSC in February 2013. 
The assembled detector was then installed into the detector vessel in Hall A of LSC. 
The assembly and installation steps are summarised below, and are presented in the pictures in \Cref{f_InstallationCanfranc4}. 
\begin{itemize}
\item Unboxing and attaching of the top PMT array to the top flange that is hanging from the crane in Hall A. 
\item Transfer of the top flange with the top array to the LSC clean room.
\item Installation of the field shaping rings and the side reflectors as well as installation of an alpha source that is dangling on a silk in the centre of the detector. The
position of the source can be adjusted in height by an external system (see \Cref{sec:3.1.1}).
\item Because of the constrained height of the clean room, 
the detector was transferred back to Hall A where it was hanged to the crane.
\item There, the fixing of the lower PMTs and final cabling was performed.
\item Finally,  the complete detector was installed inside the dewar and  evacuation was started.
\end{itemize}
Most of the time the detector was in the clean room. The total time, when parts were exposed to the normal laboratory atmosphere was kept as short as possible and was less than three hours in total. 

\subsection{Trigger electronics and DAQ}
\label{sec:2.2}

\begin{figure}[hbtp]
\begin{center}
\includegraphics[height=15pc]{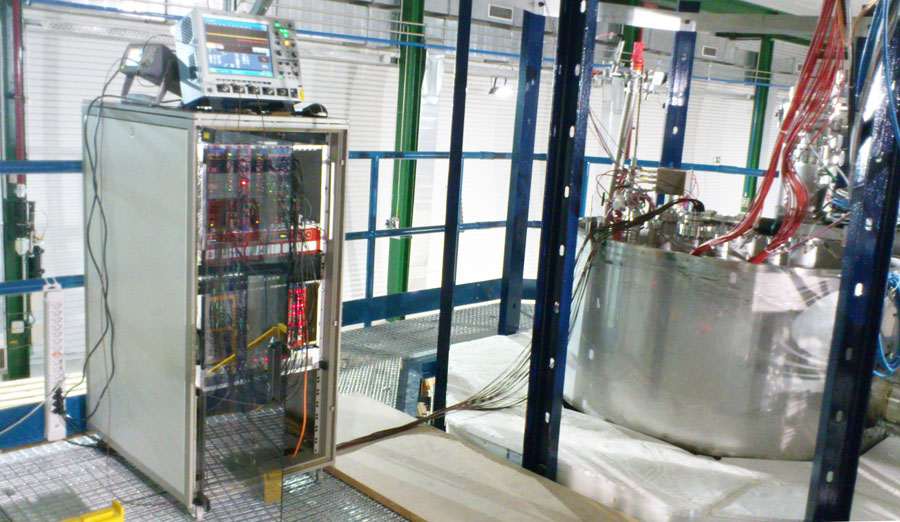}
\includegraphics[height=15pc]{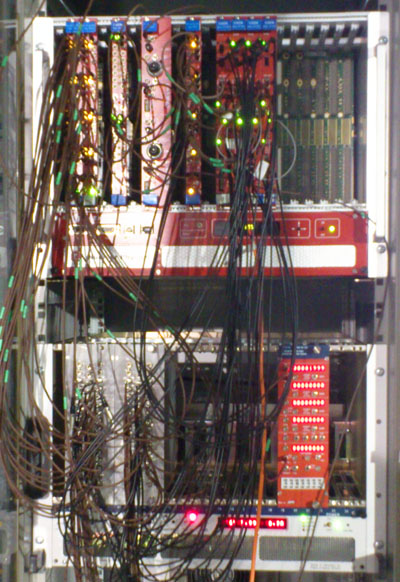}
\caption{Left: Trigger and DAQ electronics have been installed in a 
rack on the ArDM platform next to the detector vessel. Right: Close-up of the installed electronics.}
\label{fig:electronics}
\end{center}
\end{figure}

After the installation of the detector as explained in the previous section, we installed the electronics for the trigger and the data acquisition (DAQ) system in Hall A. The trigger logic was set up 
in a VME-based system 
for the data taking in gas argon, which is described below in \Cref{sec:3}. 
The 24 PMT channels are distributed over four CAEN V1720 digitizers, 
whose internal clocks are synchronised precisely so that the absolute event time (trigger time tag, TTT) can be recorded within the precision of 16 ns on each board. 
After abundantly testing the entire trigger and the DAQ chain at CERN, 
the whole system was shipped to LSC in March 2013. 
The DAQ installation and commissioning 
took place right after. 
The VME crate and an additional NIM crate, which houses a visual scaler and custom-made passive splitters used to drive each PMT signal for trigger and for recording, are installed 
on the ArDM platform, as shown in \Cref{fig:electronics}. 
Besides, a pulse generator is installed and is connected to the two LEDs installed inside the detector vessel for calibration of the PMTs. 

Following the installation, we extensively tested the entire DAQ chain. 
All of the 24 PMTs were turned on in vacuum and the LEDs were pulsed using the generator. 
The signals (mostly single photoelectrons) from all of the 24 PMT channels as well as the trigger pulses were recorded successfully with self and external (generator) triggers. 
Besides, gain curves were measured varying the PMT high voltages (HVs), using a HV control software integrated in the PLC system interface.

\section{Data taking in argon gas}
\label{sec:3}
\subsection{Measurements}
\label{sec:3.1}

First data taking at LSC was carried out for two weeks in April 2013, 
with the new detector as described in \Cref{sec:2} in warm pure gas argon, to evaluate the light yield of the new setup and the background in deep underground environment.

\subsubsection{Radioactive sources}
\label{sec:3.1.1}

For measuring the gas argon scintillation, we installed a low-activity $^{241}$Am 
source inside the detector vessel. 
The energy of the alpha's from $^{241}$Am (5.5 MeV intrinsic) provides 
(1) a well defined line (peak) for full energy deposition;
(2) an easy discrimination of low energy background/noise from alpha signals and 
(3) a relatively well localized scintillation photons emission (alpha range $\sim$4 cm in gas argon at 1 bar).
The $^{241}$Am source hence provides a straightforward way of 
studying the position dependence of the light yield in the system. 

We used a custom-made source produced at CIEMAT.
A thin $^{241}$Am foil is pressed onto a 1-mm-thick stainless steel disc and
sealed with a thin Mylar layer. 
The actual energy spectrum of the alpha's emitted through a Mylar protection layer has been measured showing negligible energy loss in Mylar. 
The total activity of the source is $\sim$500 Bq. 
The actual rate of the alpha-particles 
emitted into the gas argon is estimated to be $\sim$100 Hz, 
since half of the activity is absorbed by the stainless steel substrate, 
and the geometry of the source holder made of PTFE also 
reduces the rate and actually creates two types of ``tracks'',
as discussed later.
The source holder is hanging with a nylon thread along the axis of the cylindrical active volume, and the height of the source can be altered with the aid of a spool mounted onto a rotary motion feedthrough, over the full height of the active volume. 
The source is mounted so that alpha's are emitted only downwards. 
Alpha emission 
is accompanied by  
60-keV gamma's for 36\% of alpha decays. 
The emission rate for such gamma's is estimated to be 180 Hz 
over $4\pi$, however a large fraction of them emitted upwards is absorbed by the stainless steel substrate.

\subsubsection{Measurements in pure argon gas}
\label{sec:3.1.2}

Since the scintillation yield in gas argon is susceptible to water, oxygen or nitrogen, 
a care has to be taken to maintain a low impurity concentration of the order of 1 ppm during the measurement. 
As the gas recirculation/purification system has not yet been installed at LSC  (see \Cref{sec:8.2}), 
we observed after filling a decrease of the lifetime of the slow component of scintillation, 
consistent with an increase of impurity concentration with time, dominated by 
outgassing of the components of the detector.
However, as will be discussed later in \Cref{sec:3.2}, the time evolution of the impurity concentration 
was slow. 
This allowed us to perform a continuous measurements for typically 5--7 hours per day without refilling.  
The detector vessel thus was filled with pure gas argon every day before the measurements, and was evacuated overnight. 
The vacuum inside the vessel was as good as $5 \times 10^{-6}$ mbar before the first measurement, and  
was systematically $\sim$$2 \times 10^{-5}$ mbar before each refilling. 
A bottle of 99.9999\% pure argon gas\footnote{ALPHAGAZ 2 from AIR LIQUIDE.} was connected to the gas inlet to the main vessel via a plastic tube. After flushing the tube abundantly, the inlet valve was opened 
and the main vessel was filled up to 890 mbar, slightly higher than the ambient pressure in the lab, which was typically 870--880 mbar. 
The refilling procedure took typically 50 minutes including the flushing. 
Approximately 1500 bar$\cdot$L of gas argon was spent for each filling, which allowed 6--7 fillings per standard gas bottle. 

\subsubsection{PMT calibration}
\label{sec:3.1.3}

\begin{figure}[hbtp]
\begin{center}
\includegraphics[height=15pc]{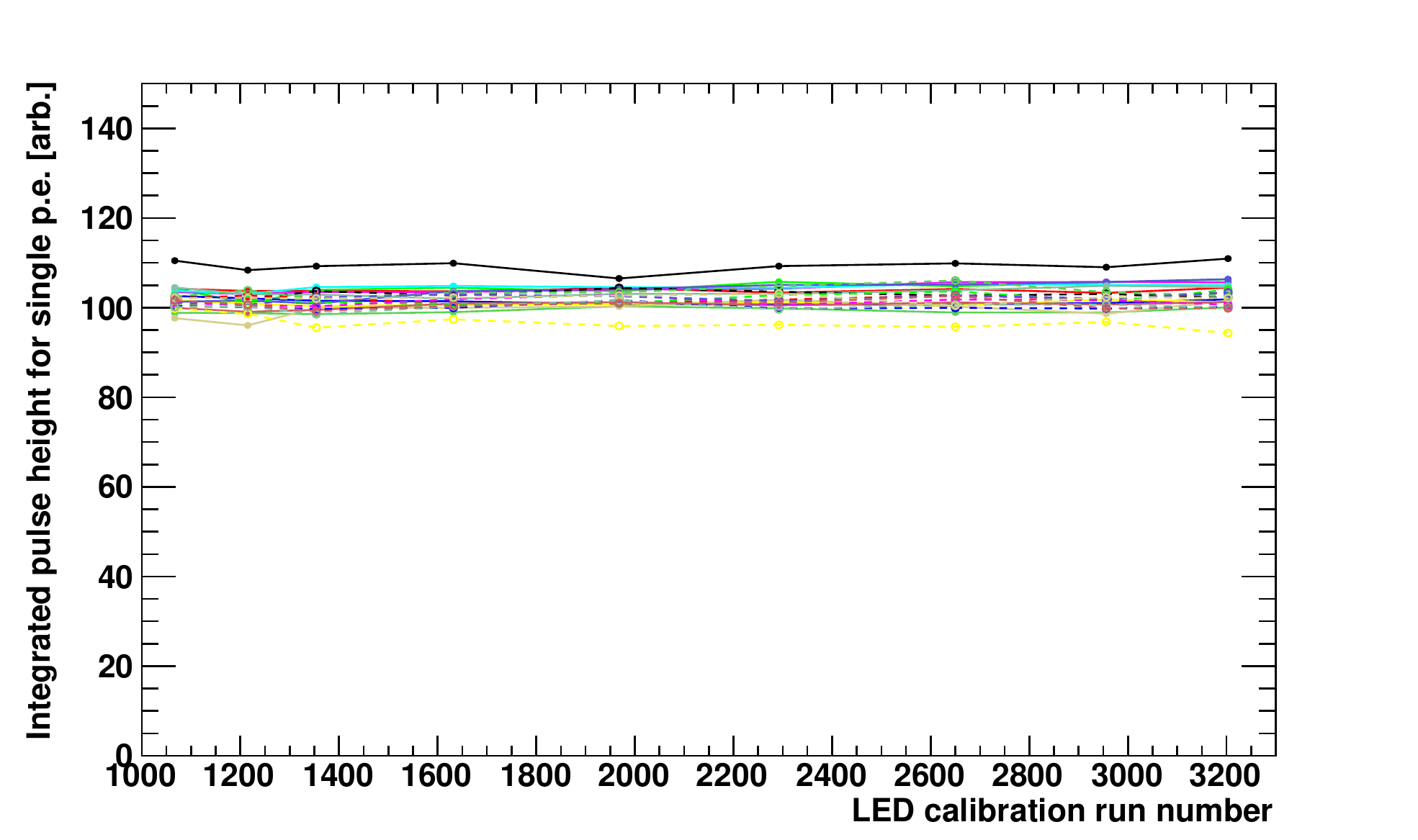}
\caption{Mean value of the integrated pulse height for single photoelectron pulses recorded by 23 PMTs during calibration runs using LED. Calibration runs were taken once every day during the test.}
\label{fig:pmtresponse}
\end{center}
\end{figure}

The operation HV for each PMT was determined before the
test, based on the gain curves measured during the electronics installation as described above in \Cref{sec:2.2}. 
The HVs ranging from 976 V to 1333 V among the PMTs are chosen to obtain the gain of 
$5 \times 10^7$ from each PMT, resulting in the 
peak pulse height of $\sim$20 mV after the passive splitter for single photoelectrons. 
\Cref{fig:pmtresponse} shows the mean value of the integrated pulse height for single photoelectron pulses, measured for 23 PMTs during calibration runs using LED, on each of nine days of the test. 
We discovered a problem on one PMT, i.e. PMT18 located at the edge of the bottom array, which stopped operating during this test due to 
a loose contact between the PMT and its base. 
This problem has been fixed 
after the test by re-opening the detector. 
The gain was consistent throughout the test period within $<$5\%. 
The variation of the gain among PMTs was within $\pm$5\% except for one PMT which showed systematically 
higher gain by 10\% compared to the others. 
It is worthwhile to note that we have not observed any spark of the PMT HVs, even though the pressure of gas argon was less than 1 bar, demonstrating the good design of the new base, which was done carefully to minimise risks for sparks. 
As the nominal operation voltage of this type of PMT 
is 1700 V, we still have a large room to increase the operation HVs if necessary. 

\subsubsection{Trigger}
\label{sec:3.1.4}

The trigger logic was optimised for the test in gas argon. 
Briefly, the trigger is generated in the following manner; 
(1) The analog signals from 12 PMTs of each of the top and the bottom array are summed up after the passive splitter, using an analog fan-in/out. Note that one of the bottom PMTs, i.e. PMT18 was not providing signals as described above. 
(2) The analog sum of each array is discriminated using a discriminator. 
(3) AND of the two discriminator outputs (each 70 ns wide) is made using a coincidence module 
that finally generates the trigger signal. 

Most of the runs were taken with the discriminator threshold of 35 mV for each of the two analog sums. 
This threshold voltage corresponds to $\sim$1.75 p.e. (photoelectrons) for each array and to $\sim$3.5 p.e. in total for the coincidence, producing a raw trigger rate at $\sim$100~Hz. 
Some data were taken with 100-mV threshold corresponding at least to 5 p.e. per array and to 10 p.e. in total, resulting in the trigger rate of typically 50 Hz. 
Trigger without coincidence, using the discriminator output for the bottom array only, was employed with pre-scaling down to 400 Hz for investigating threshold effects. 
In addition we took data with trigger generated by the pulse generator at 200-Hz repetition rate to study the dark count rate of the PMTs.

\subsection{Analysis}
\label{sec:3.2}


%

\subsubsection{Signal of the alpha's from the internal $^{241}$Am source}
\label{sec:3.2.2}

\begin{figure}[hbtp]
\begin{center}
\includegraphics[height=20pc]{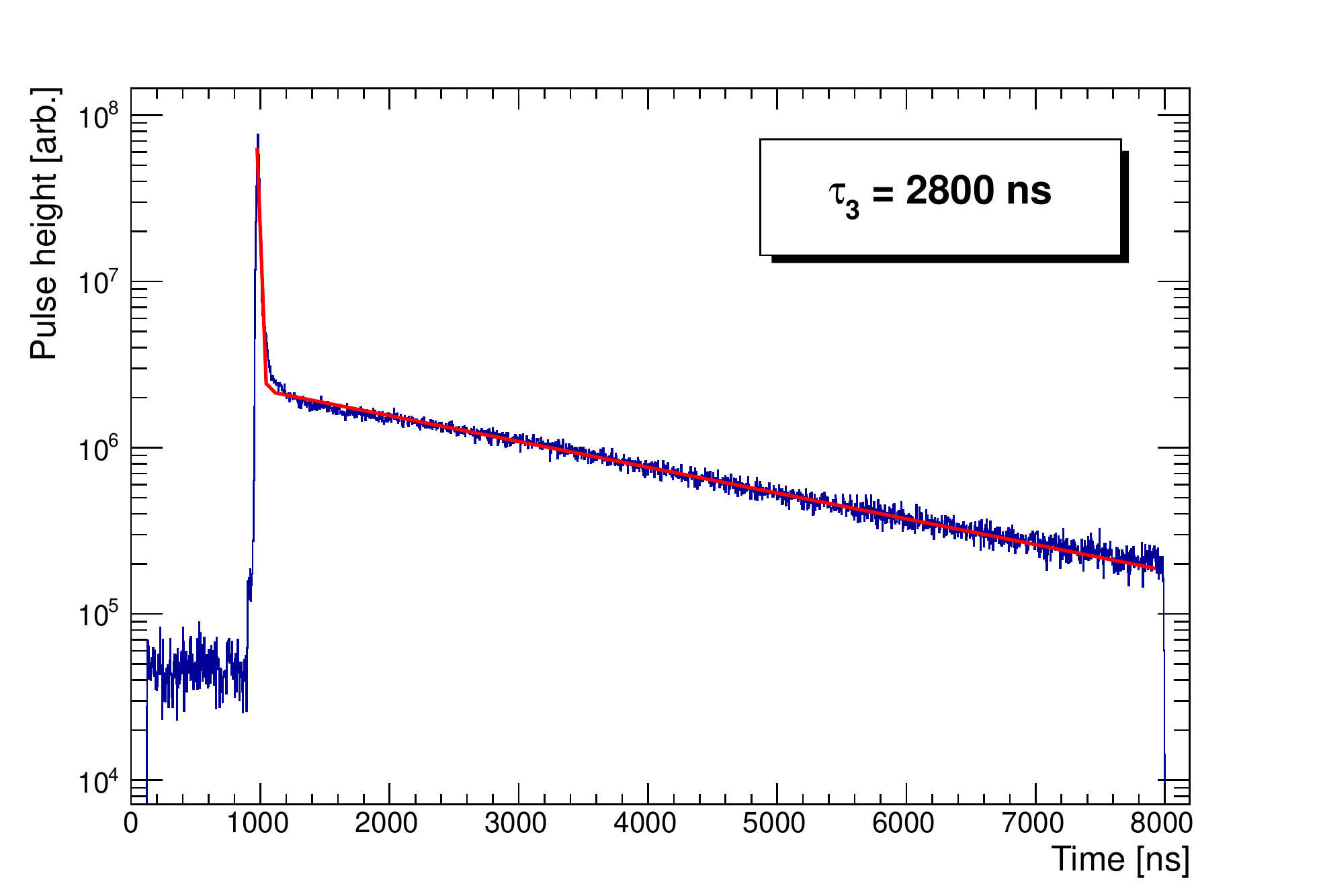}
\caption{Averaged waveform for the measured alpha events from the internal $^{241}$Am source. The red curve shows the fit with a function involving two exponential functions corresponding to the decays of the fast and slow components.}
\label{fig:meantrace}
\end{center}
\end{figure}

\Cref{fig:meantrace} shows an averaged waveform, 
showing the characteristic signatures of the argon scintillation signals, 
consisting of fast and slow components, due to the radiative decays of the 
singlet and triplet states, respectively, of the argon excimer to the dissociative ground state. 
The waveform can be fitted with a function 
involving two exponential functions corresponding to the two components. The decay time of the slow component, denoted as $\tau_3$ hereafter, and consequently the number of detected photoelectrons in it, depends on the purity of argon in the region around 1 ppm in impurity concentration, while the fast component essentially is unaffected. 
Below 0.1~ppm the dependency becomes weak and $\tau_3$ ultimately reaches $\sim$3200 ns at 1 ppb level (see \cite{Boccone:2009kk} and its references). 

In the following analysis we define the light yield ($LY$) as the number of detected photoelectrons contained in both the fast and the slow components, i.e. the integral of each event's waveform over the full time range
divided by the corresponding integral for a single photoelectron pulse. 
We calculate the $LY$ for the top array ($LY_{\rm top}$) and for the bottom one ($LY_{\rm bottom}$), 
as well as for the sum ($LY_{\rm TOT} = LY_{\rm top} + LY_{\rm bottom}$). 
It should be noted that the results discussed below were obtained with 22 PMTs, 11 each of the top and bottom arrays, as the problematic PMT8 (high dark counts) and PMT18 (not working) were excluded from the whole analysis. 
PMT8 however was included in the trigger. 

\begin{figure}[hbtp]
\begin{center}
\includegraphics[height=12.5pc]{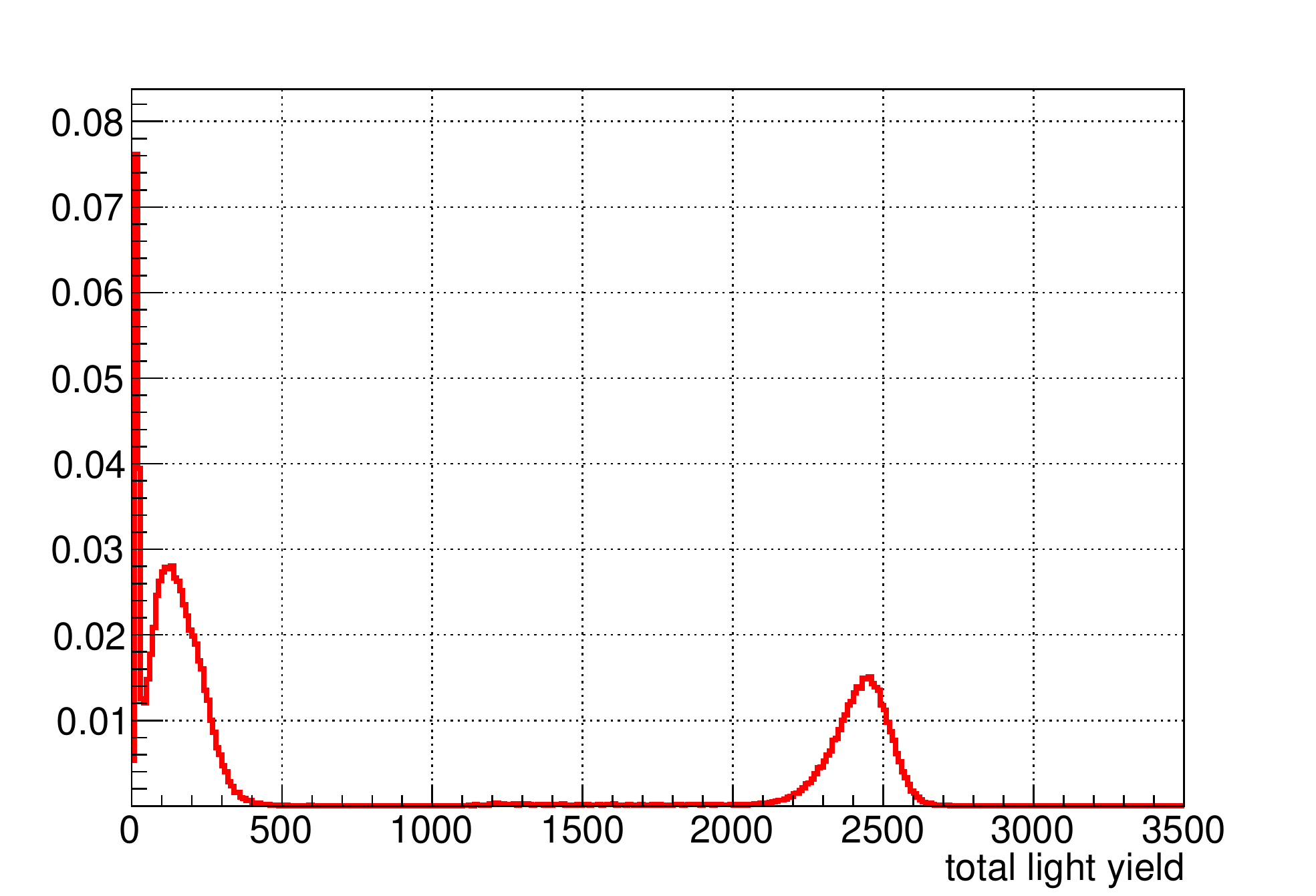}
\includegraphics[height=12.5pc]{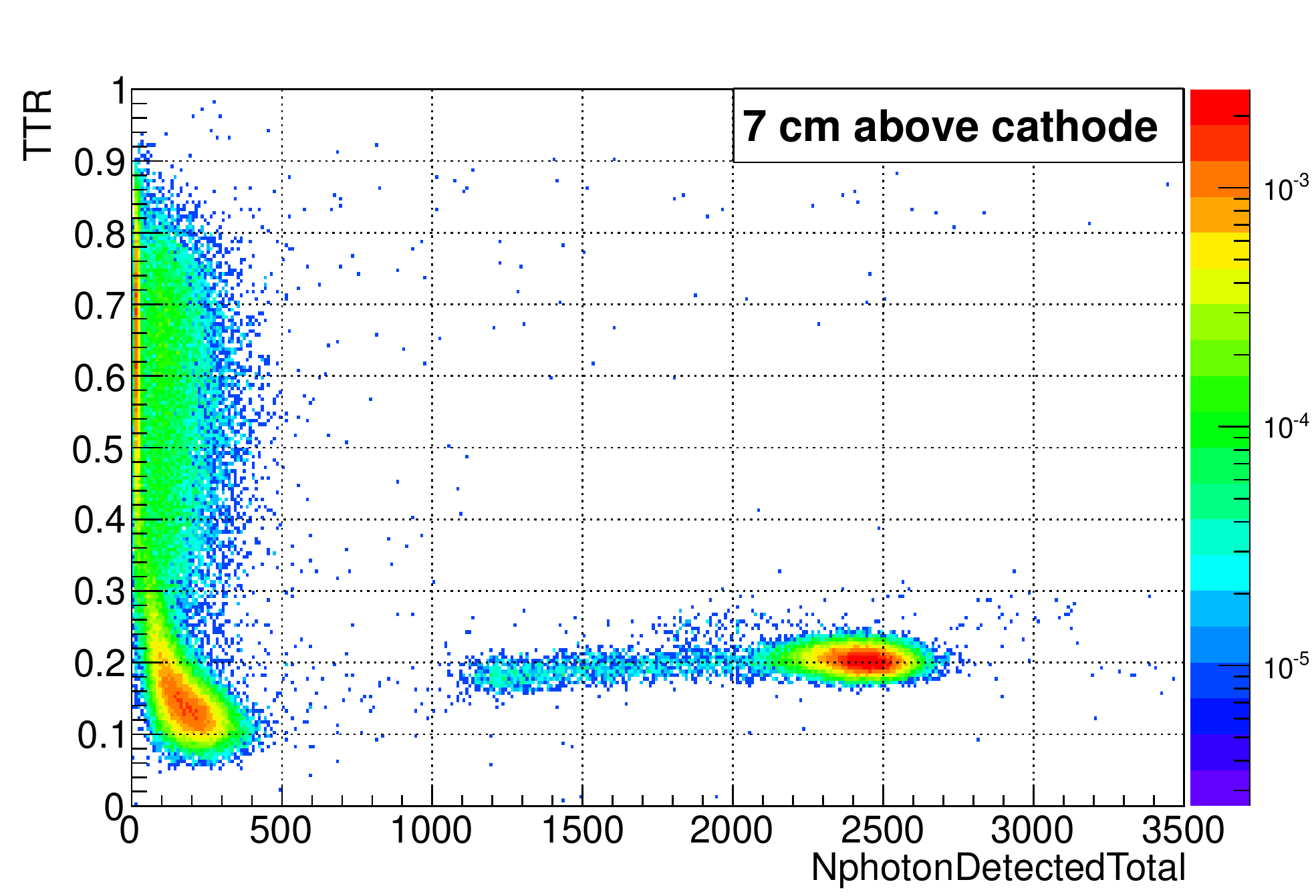}
\caption{Left: Total light yield  in p.e. ($LY_{\rm TOT}$) 
measured with the $^{241}$Am source at 7 cm above the cathode. Right: Top-to-total ratio ($TTR$) as a function of $LY_{\rm TOT}$ for the same run as in the left plot. See text for the definition of the variables.}
\label{fig:IPH}
\end{center}
\end{figure}

\begin{figure}[hbtp]
\begin{center}
\includegraphics[width=\columnwidth]{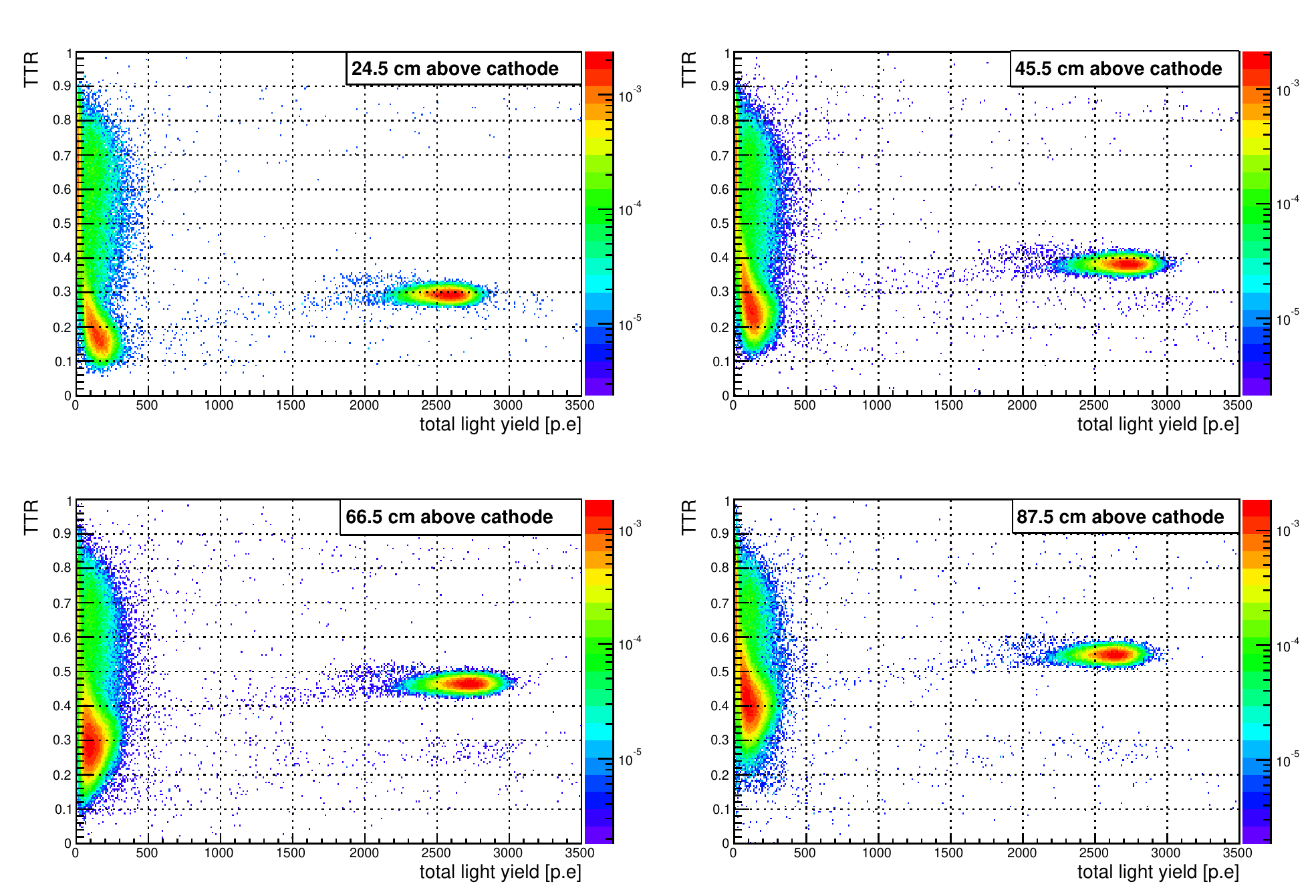}
\caption{Top-to-total ratio ($TTR$) is plotted as a function of total light yield for different positions of the internal $^{241}$Am source. The two red islands move upwards together with the source position, indicating that those events are due to the alpha's from the source.}
\label{fig:movingislands}
\end{center}
\end{figure}

The left plot of \Cref{fig:IPH} shows the distribution of the event light yield for a typical run, where the $^{241}$Am source was positioned close to the bottom PMT array (7 cm above the cathode grid). 
The high energy peak seen at around 2500 p.e. corresponds to the situation
where the alpha's deposit the full energy in the gas argon. 
To understand the nature of the low energy peak below 500 p.e. we calculated the top-to-total ratio $TTR = LY_{\rm top}/LY_{\rm TOT}$ and plotted it as a function of $LY_{\rm TOT}$ in the right plot of \Cref{fig:IPH}.
In this plot, we clearly see three different types of events:
\begin{enumerate}[(1)]
\item The ``golden'' alpha events: the red island at $TTR \sim 0.2$ and $LY_{\rm TOT} \sim 2500$ p.e., which are due to the alpha's depositing their full energy in the gas argon. 
\item The alpha ``short track'' events: the red island at $TTR < 0.3$ and $LY_{\rm TOT} < 500$ p.e., which are due to the alpha's hitting the source holder (serving like a collimator) soon after the emission, depositing only a small fraction of the energy in the gas argon. 
\item Background events: the green cloud visible at $TTR > 0.3$ and $LY_{\rm TOT} < 500$ p.e, where the signals are seen equally by the top and the bottom arrays, indicating the events are distributed all over the active volume. 
\end{enumerate}
Since the $^{241}$Am source was positioned close to the bottom PMT array and the source is always pointing downwards, it is reasonable that for this run the scintillation photons of all the alpha-induced events fall preferably onto the bottom array, resulting in small $TTR$. 
This interpretation can be confirmed by the fact that the two red islands move together depending on the $^{241}$Am source position, as shown in \Cref{fig:movingislands}. 

Apart from the three event types as described above, it is interesting to mention that two faint bands are seen in the right plot of \Cref{fig:IPH} and in \Cref{fig:movingislands}, at $TTR \sim 0.8$ and also at $TTR \sim 0.3$ up to the high energy region. These faint bands are independent of the position of the $^{241}$Am source.
They may be due to the alpha's emitted from the PMT arrays, although it still needs to be clarified. 
The counting rate of those events is estimated to be roughly of the order of 0.05 Hz. 

\begin{figure}[hbtp]
\begin{center}
\includegraphics[height=15pc]{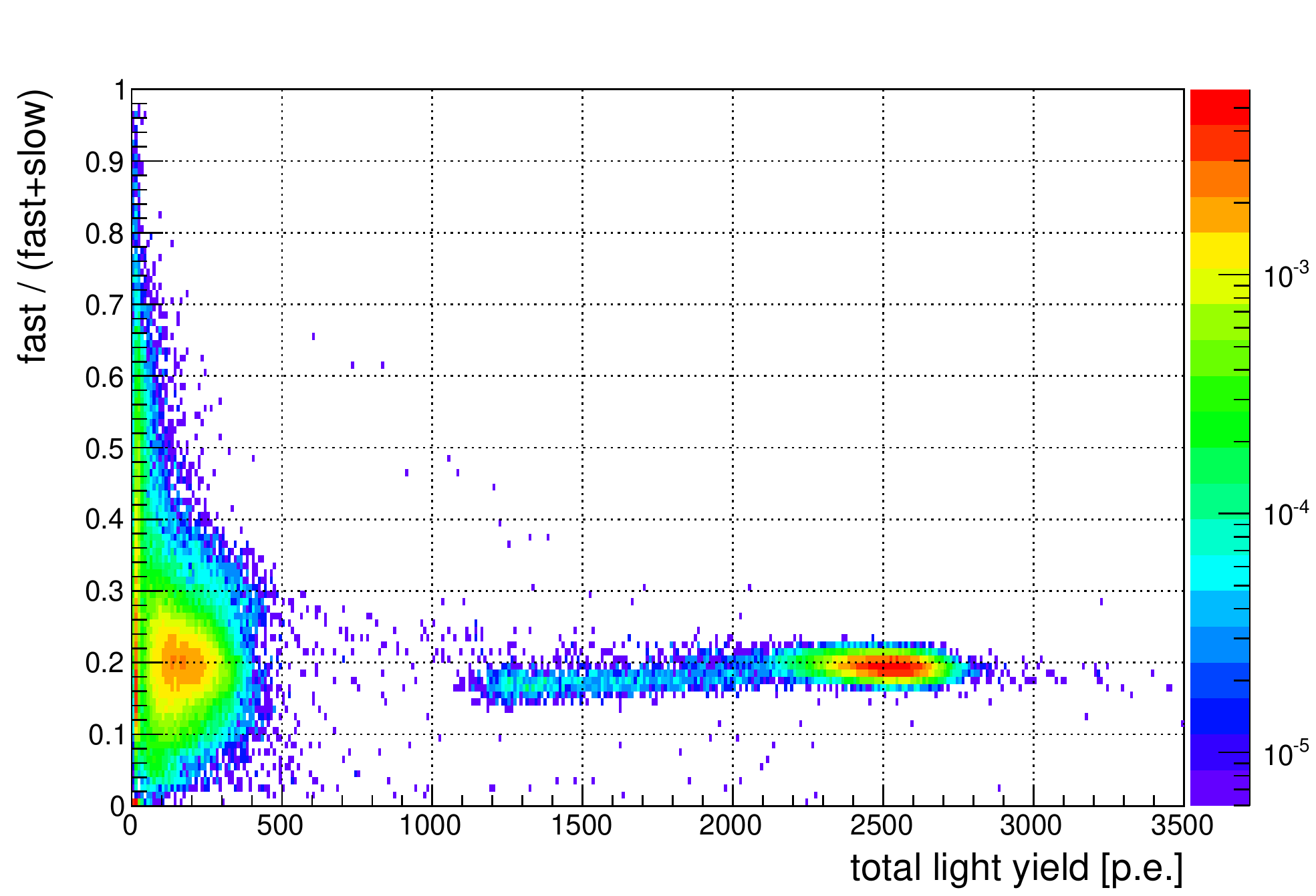}
\caption{Component ratio ($CR$) plotted as a function of the total light yield in p.e. 
for the same run as for \Cref{fig:IPH}.}
\label{fig:CR}
\end{center}
\end{figure}

The component ratio ($CR$) is defined as the ratio of the number of detected 
photoelectrons in the fast component to the total light yield. 
In \Cref{fig:CR} the $CR$ ratio is plotted as a function of the total light yield, for the same data set as used in \Cref{fig:IPH}. 
Such a parameter will be used for the pulse shape discrimination (PSD) between the electron and the nuclear recoil events, and is an essential tool for the WIMP search in ArDM. 
In  argon gas at $\sim$1~bar, the alpha particles show similar $CR$ as electron recoils (Ref. \cite{Boccone:2009kk} and its references).
As can be seen in \Cref{fig:CR} essentially all the events are populated in the band around $CR \sim 0.2$ as expected. 
At lower energy the most probable value stays the same although the band becomes more spread, affected 
by photoelectron statistics. 

\begin{figure}[hbtp]
\begin{center}
\includegraphics[height=12pc]{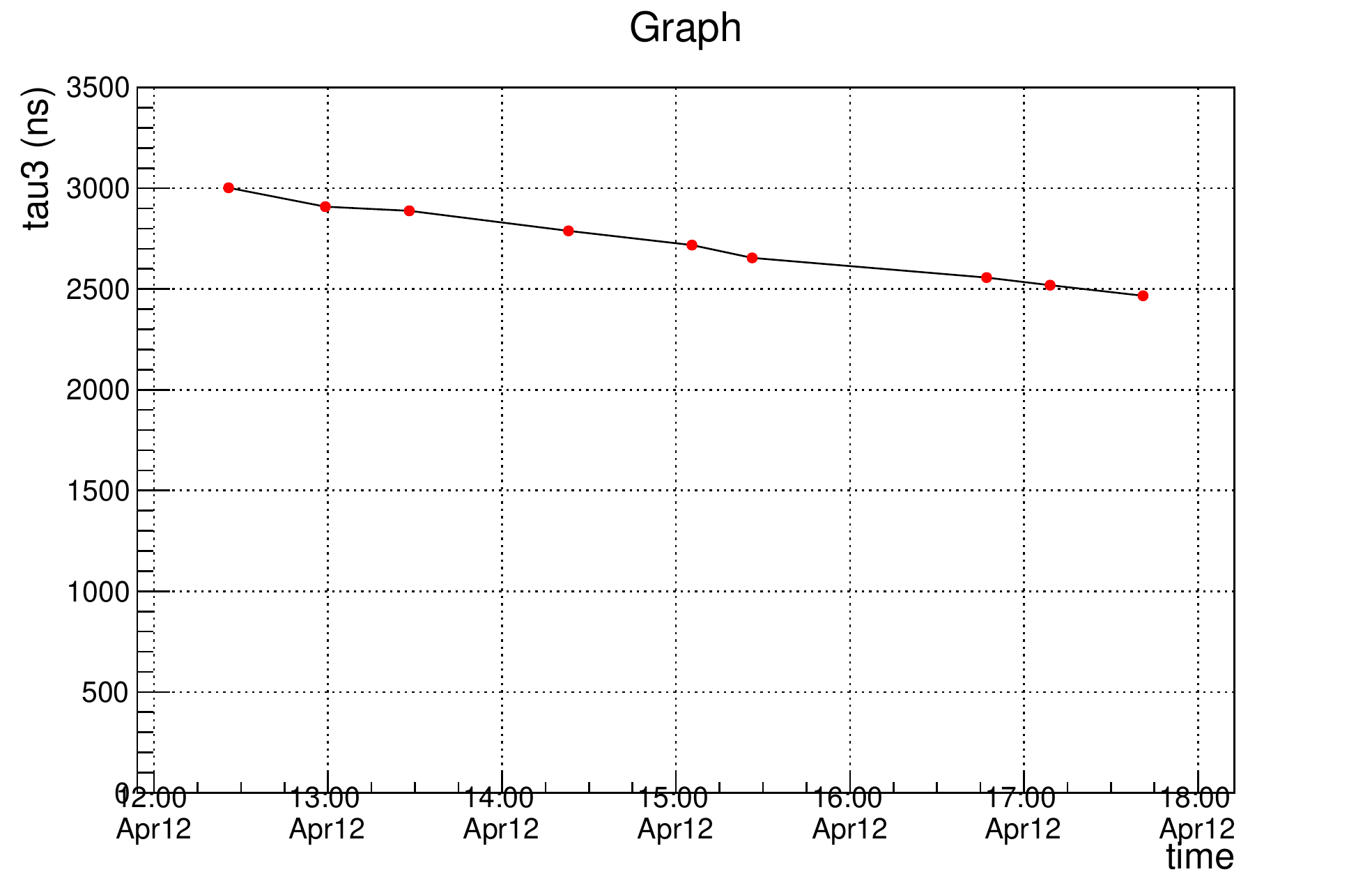}
\includegraphics[height=12pc]{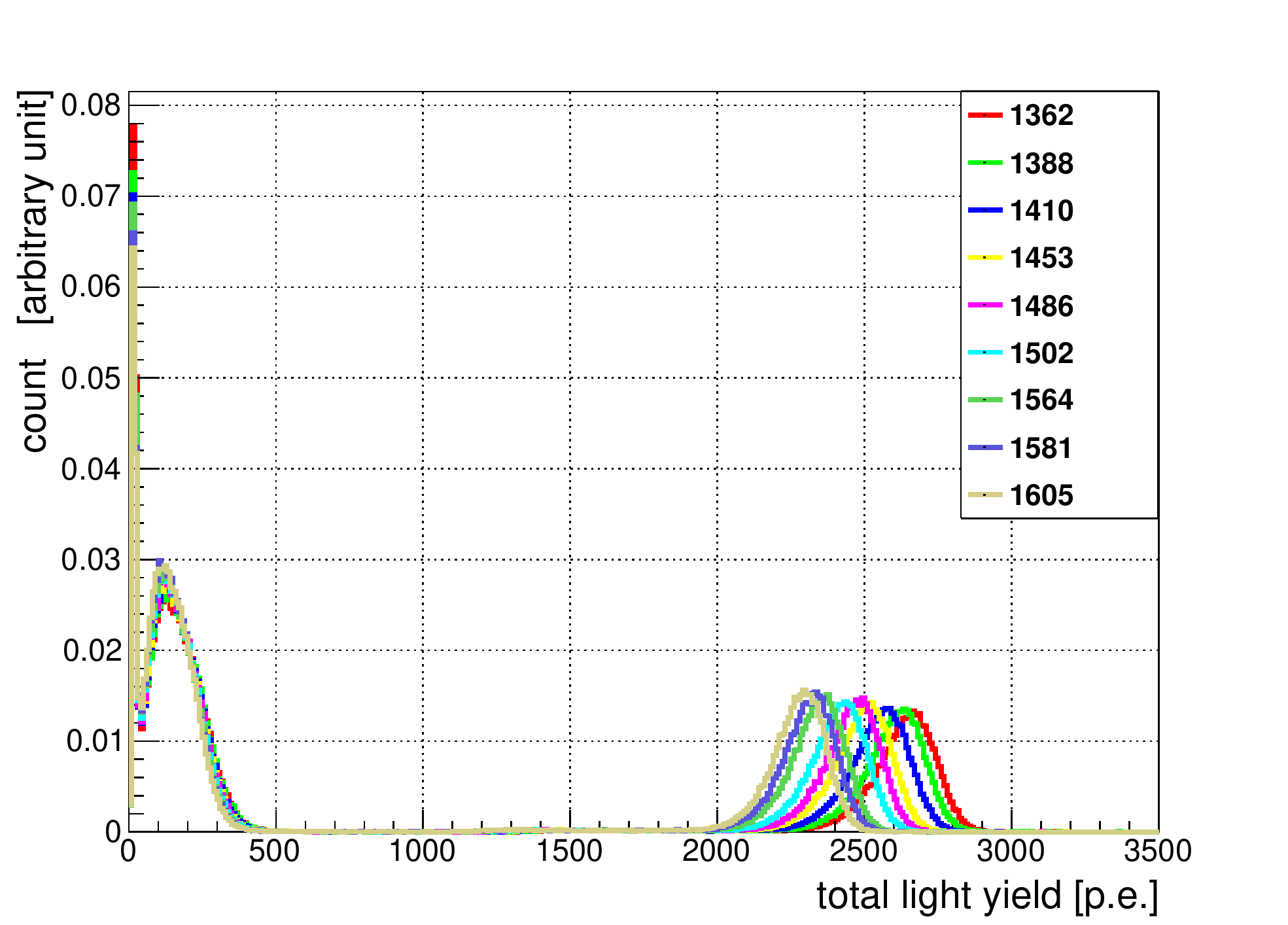}
\caption{Left: Evolution of $\tau_3$ over five hours due to outgassing from the detector components. Right: Corresponding evolution of the total light yield.}
\label{fig:tau3evolution}
\end{center}
\end{figure}

To assess the light yield in the new light readout system, we focus on the ``golden'' events in category (1). 
We took data with different positions of the internal $^{241}$Am source 
along the axis of the cylindrical active volume of ArDM, to investigate position dependence of the light yield in the system. 
First, we applied a correction on the measured light yield for gas argon purity.
\Cref{fig:tau3evolution} shows the evolution of $\tau_3$ during five hours of measurement, where the position of the $^{241}$Am source stayed unchanged. 
It was approximately 3000 ns at the beginning and went down to 2450 ns after five hours. 
Although the change in $\tau_3$ was not large, the measured light yield also changed with time, as can be seen for the ``golden'' alpha peak in the right plot. The reduction of the total light yield by the end of this single
run was about 20\%.

\begin{figure}[hbtp]
\begin{center}
\includegraphics[height=15pc]{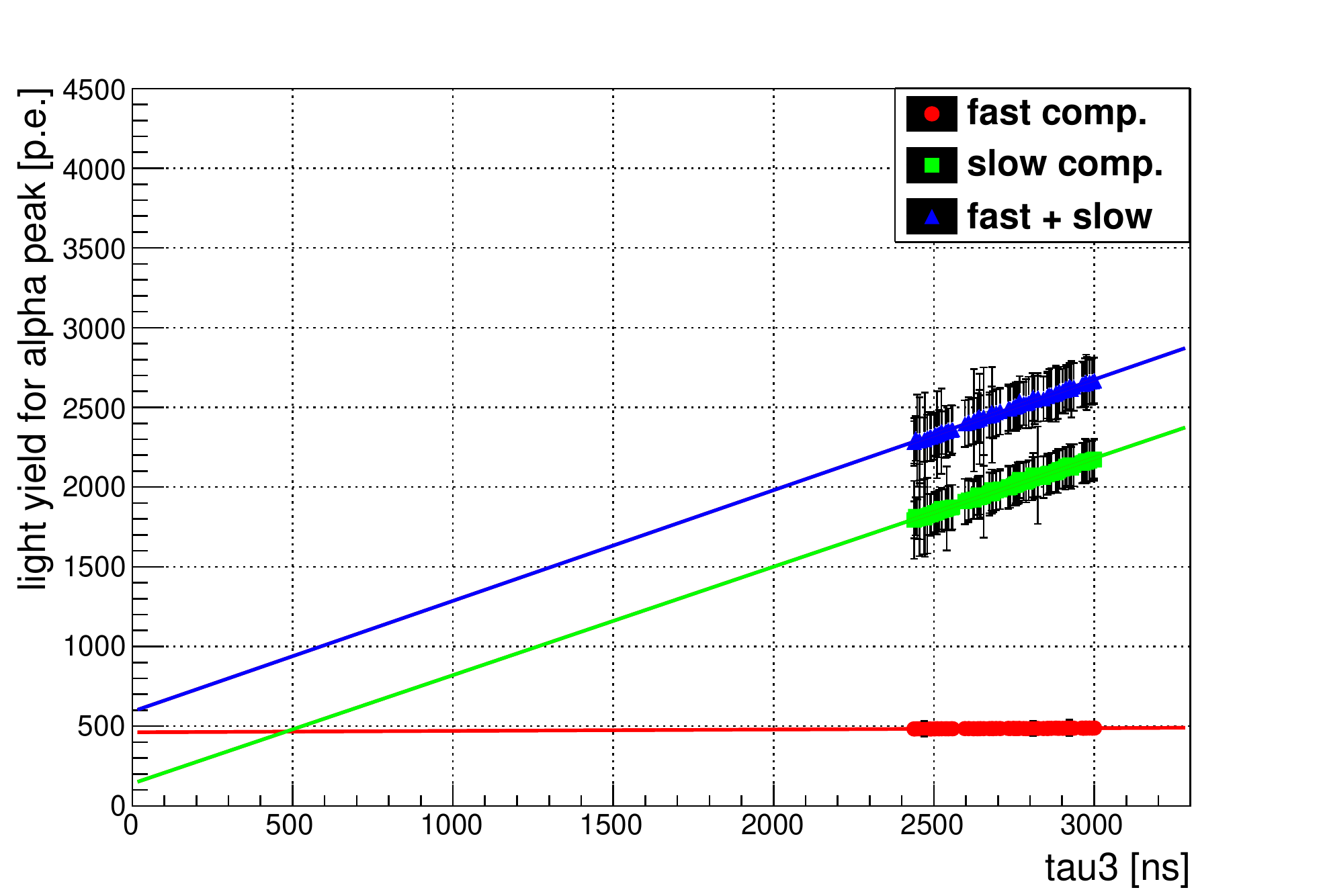}
\caption{The measured total light yield (blue) and its fast (red) and slow (green) component for the alpha peak plotted as a function of the decay time for the slow component $\tau_3$.} 
\label{fig:purity_correction}
\end{center}
\end{figure}

\begin{figure}[hbtp]
\begin{center}
\includegraphics[height=15pc]{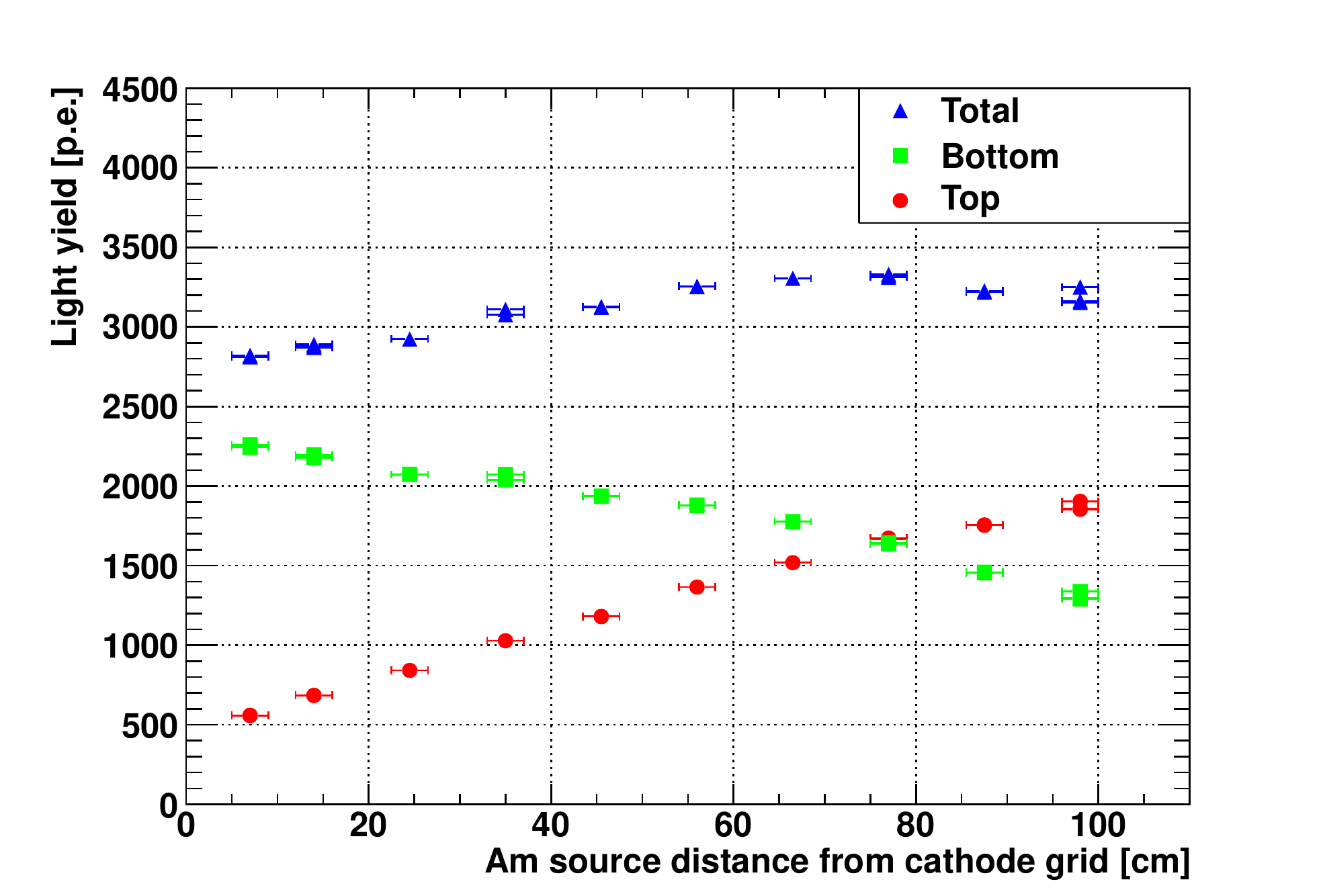}
\caption{The measured light yield extrapolated to $\tau_3 = 3200$ ns for the top (red), the bottom (green) PMT array and for the sum (blue), plotted as a function of the distance between the cathode grid and the $^{241}$Am source position.}
\label{fig:positiondependence}
\end{center}
\end{figure}

The total light yield and that of the slow component depend linearly on $\tau_3$, 
while the fast component is constant to first approximation, as shown in \Cref{fig:purity_correction}. 
In order to compare different measurements,
the total light yield is rescaled for an equivalent decay time of $\tau'_3 = 3200$ ns, corresponding
to an essentially pure argon situation.
We do a simple extrapolation: 
for each condition, from the measured $\tau_3$, $LY^{\rm TOT}$ and $LY^{\rm fast}$ the linear function can be approximated as: 
\[
LY^{\rm TOT}(\tau_3^{\prime}) = LY^{\rm fast} + \frac{LY^{\rm TOT}(\tau_3) - LY^{\rm fast}}{\tau_3}\cdot \tau_3^{\prime}\,.
\]
This function was used for the purity correction in \Cref{fig:positiondependence}. 

\Cref{fig:positiondependence} shows the light yield as a function of the distance between the cathode grid and the $^{241}$Am source. 
The light yields obtained from individual arrays show a clear dependence on the source position, 
while the dependency is much weaker when the two arrays are summed up, and the total light yield above 3000 p.e. was obtained consistently at the positions 35 cm or higher. 
It should be noted that the emission of UV photons from the alpha tracks is supposed to have a strong  top/bottom asymmetry in this configuration of the $^{241}$Am source and the holder structure.  
This effect can be seen by the fact that overall the bottom array sees more light, as naively expected. 
Therefore this data has to be understood better with the aid of the full Monte Carlo simulation of ArDM, which is an important next step for analysis. 
Nonetheless, for the measured data we obtained a good uniformity, which was $\pm$10\% over the full height of the ArDM active volume. 

\subsubsection{Improvement in light yield of the new light readout system}
\label{sec:3.2.3}

The results presented above demonstrate an important improvement in light yield of the new light readout system compared to the prototype one used in previous measurements on surface. 
In a similar measurement done at CERN with the prototype system using a movable $^{241}$Am source in gas argon, the obtained total light yield extrapolated to $\tau_3 = 3200$ ns was at most 850 p.e. with a strong dependence on the source position. 
In this test we obtain $\sim$3000 p.e. for the position of the alpha peak, consistently over a large fraction of the full height of the active volume. 
At CERN we used a palladium-sealed  $^{241}$Am source, which was measured to be peaked at 4.8 MeV.
The actual energy of  the present source, sealed with a Mylar layer, has been measured and is close to 5.5 MeV, which is $\sim$15\% higher than the other. 
Taking this difference into account, the gain in the  light yield is conservatively larger than a factor of 3.

The light yield measured with the old system in LAr using 511 keV gamma's from $^{22}$Na source was $\sim$0.7 p.e./keVee with zero electric field \cite{Degunda:2013Diss,Lazzaro:2012Diss}. 
Assuming that the improvement in light yield will be unchanged in LAr, 
simply multiplying the factor of 3 as obtained above to 0.7,  
we expect at least 2 p.e./keVee for the new system in LAr with zero electric field.

\subsubsection{Background events}
\label{sec:3.2.4}

\begin{figure}[hbtp]
\begin{center}
\includegraphics[height=12pc]{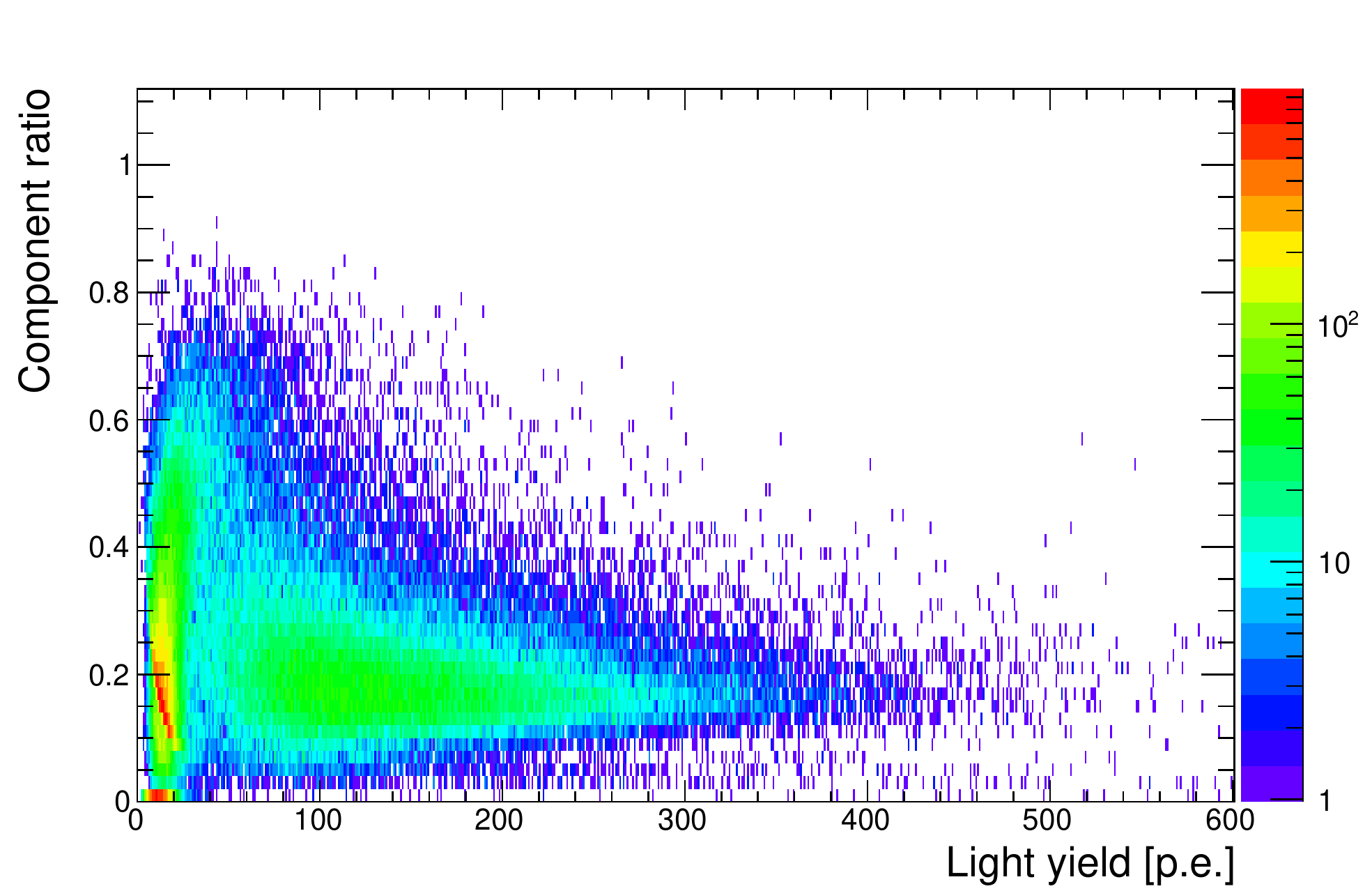}
\includegraphics[height=12pc]{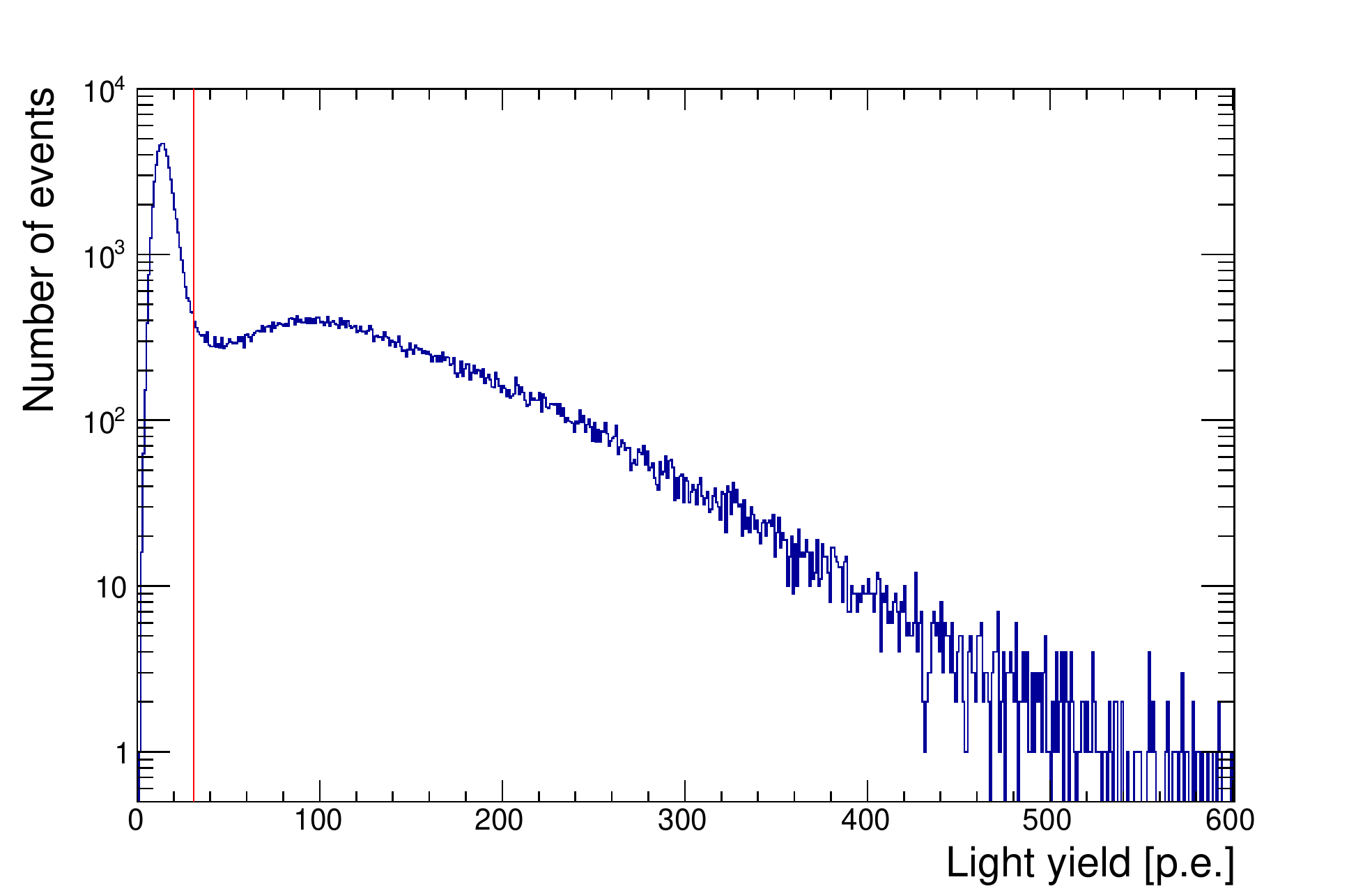}
\caption{Left: Component ratio ($CR$) plotted as a function of the total light yield ($LY_{\rm TOT}$) for the events $TTR>0.4$ for the same run as in \Cref{fig:IPH}. Right: Projection of the left plot to the horizontal axis. The red vertical line corresponds to the energy threshold determined by dark rate conditions (see text).}
\label{fig:gammalike}
\end{center}
\end{figure}

To look into  low energy events not consistent with the alpha's from the internal $^{241}$Am source, we selected the events requiring $TTR > 0.4$ for the same run as in \Cref{fig:IPH}.
The component ratio for those events (the events type 3 as described in \Cref{sec:3.2.2}) is plotted as a function of the total light yield in the left plot of 
\Cref{fig:gammalike}. 
Above $\sim$30 p.e. those events also mostly are populated around $CR = 0.2$, indicating that they are consistent with physical argon scintillation signal. 
The band becomes broader with decreasing total light yield due to lower photoelectron statistics. 
The projection of the left plot of \Cref{fig:gammalike} to the horizontal axis is shown in the right. 

The narrow peak centred at $LY_{\rm TOT} = 14$ p.e. is accounted as the contribution from noise events due to the dark counts from the PMTs. The event rate for this type of events is $\sim$13~Hz. 
Analysing the data taken with random trigger, we obtained the dark count rate as follows; 
(1) We had one noisy PMT8 located at the edge of the top PMT array, which on average were giving 13 dark counts in the 8-$\mu$s acquisition window. This PMT was removed from all the analysis, while it was included in the trigger. 
(2) The other 11 PMTs of the top array were giving total of 8.9 dark counts on average. 
(3) The 11 PMTs of the bottom array (apart from PMT18 which was not giving signal) were giving total of 2.3 dark counts on average. 
(4) Consequently 11.2 dark counts are recorded from the total of 22 PMTs used in the analysis. 
A crude estimation of the rate of random coincidences of the dark counts which can generate the trigger signal agrees with the order of 10 Hz. 

The continuum up to $\sim$600 p.e. is considered as physical events due to the scintillation of gas argon, induced by ambient and material radiation. 
As already mentioned in \Cref{sec:3.1.4}, the trigger logic requires $\sim$2 p.e. for the analog sum of each of the top and the bottom array. 
Requiring AND of the two arrays (within $\sim$140 ns) $\sim$4 p.e. are needed in the fast component to issue a trigger, corresponding to $\sim$20 p.e. in total light yield as the mean value of the component ratio is 0.2 (i.e. $4/0.2 = 20$). 
For the recorded number of photoelectrons, this number is pushed up to 31 due to the average dark counts from the 22 PMTs. This value is shown with a red vertical line in the plot. 

Above the red line, the event rate is calculated to be $\sim$13 Hz. 
The individual 
origin of those physical background events is to be understood, as well as the bump at around 100 p.e.. 
Sources for such events are: (1) background gamma (or beta) radiations from the environment and inner detector components, (2) 60 keV gamma's from the internal $^{241}$Am source and (3) beta decays of $^{39}$Ar.  The event rate from (2) and (3) however are estimated to be of the order of 1 Hz each. 



\section{Safety analysis and quantitative risk assessment	}
\label{sec:4}

For the underground operation of ArDM at LSC, which involves total of $\sim$2 tons of cryogenic LAr, 
we are now performing a Quantitative Risk Assessment (QRA)\footnote{In collaboration with 
Institute of Nuclear \& Radiological Sciences \& Technology Energy \& Safety,
National Center for Scientific Research ``DEMOKRITOS'', Athens, Greece.}. This consists of:
\begin{enumerate} 
\item \textbf{Qualitative Safety Analysis:} Identification of immediate causes of Loss of Containment (LOC) resulting in argon releases during any of the operational phases, and the initiating events (IE) as well as the safety measures implemented to impede the IEs from resulting in a LOC. Delineation of accident sequences using appropriate Event \& Fault Tree models. Binning of the accident sequences in plant damage states and release categories. 

\item \textbf{Consequence Assessment:} Calculation of argon dispersion for each of the identified release categories coupled with possible states of available mitigating measures and degree of implementation of emergency response.

\item \textbf{Quantitative Risk Assessment:} Estimation of accident probabilities along with consequence probabilities and generation of a complete risk registry and risk matrix.

\item \textbf{Evaluation of Safety Management System:} Evaluation of  the safety systems implemented on site including organisation, emergency plans and safety related maintenance activities.

\item \textbf{Evaluation of additional safety measures:} Evaluation of potentially additional safety measures in collaboration with those responsible for the experimental facility.
\end{enumerate}

\section{Systematic material screening}
\label{sec:5}

The material screening campaign is ongoing. 
One PMT, its support structure, one fully assembled PMT base, field cage resistors and parts of the PMT supporting structure have been screened. 

\begin{figure}[hbtp]
\begin{center}
\includegraphics[height=15pc]{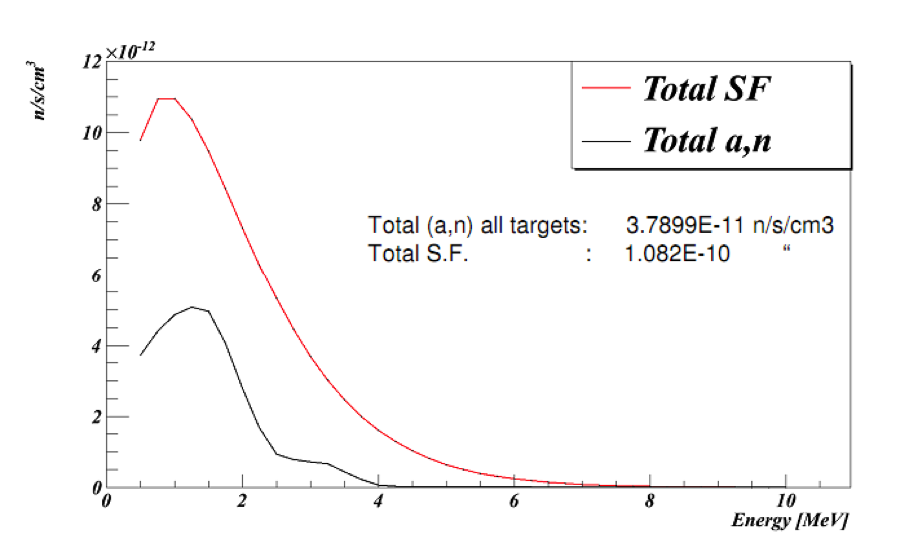}
\caption{Preliminary results of the estimation of the flux and spectrum of the neutrons coming from the material contamination. Spontaneous fission and ($\alpha$, $n$) neutrons from 1 ppb of $^{238}$U in the stainless steel are shown in red and black lines, respectively. The code SOURCES4C has been used.}
\label{fig:screening}
\end{center}
\end{figure}

Together with the first measurements of the material contamination we performed 
the simulations necessary to evaluate the background given by fission and ($\alpha$,$n$) neutrons. The code SOURCES4C has been used to estimate the flux and the spectrum of the neutrons coming from the material contamination. 
Preliminary results of such simulations for the assumed 1-ppb contamination of $^{238}$U in the stainless steel are shown in \Cref{fig:screening}. 
The spectra thus obtained can be scaled according to the results of the material screening, to eventually evaluate the neutron flux in the actual setup. 

\section{Temperature and radon monitoring}
\label{sec:6}

\begin{figure}[hbtp]
\begin{center}
\includegraphics[height=12.5pc]{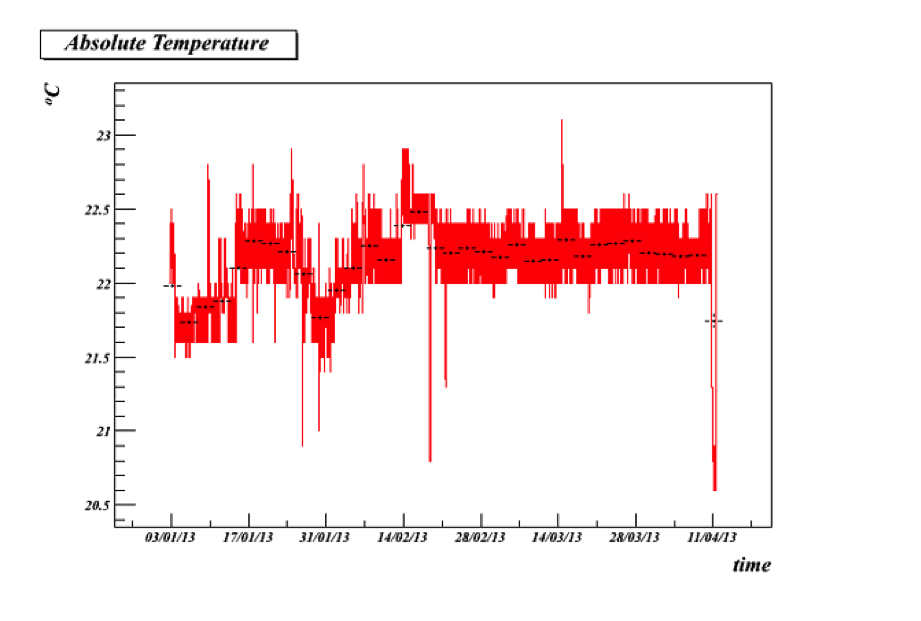}
\includegraphics[height=12.5pc]{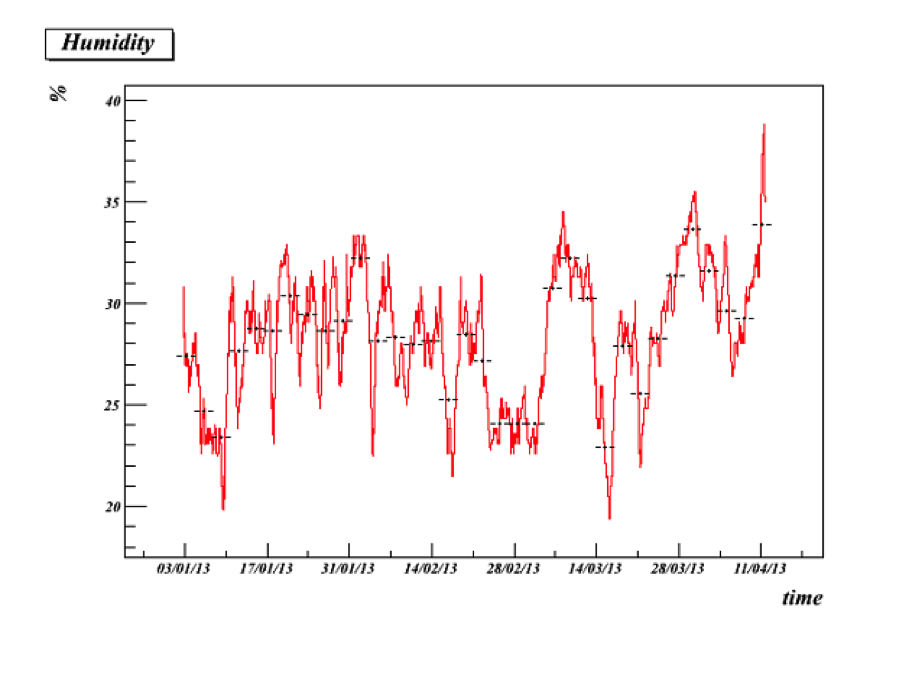}
\includegraphics[height=12.5pc]{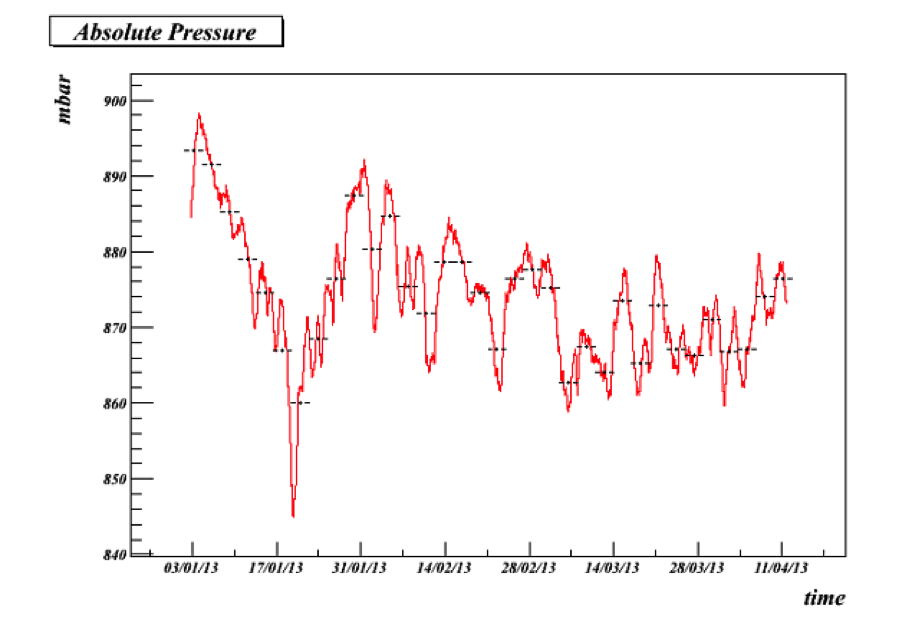}
\includegraphics[height=12.5pc]{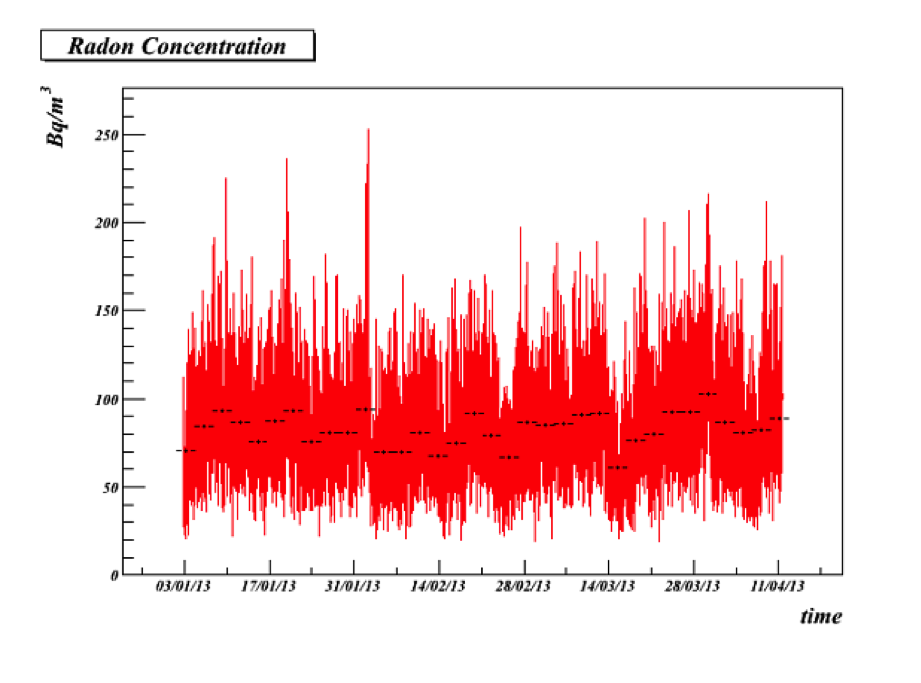}
\caption{Environmental parameters in Hall A measured since the beginning of the year 2013 with the monitoring station of the lab placed in proximity of the ArDM detector. Plotted are: temperature (top left), humidity (top right), pressure (bottom left) and radon concentration (bottom right).}
\label{fig:env}
\end{center}
\end{figure}

We are monitoring several environmental parameters in 
Hall A. 
We started the analysis of the data collected since the beginning of 2013 with the monitoring station placed in proximity of the ArDM detector. 
The measured temperature, humidity, pressure, and radon concentration in air are presented in \Cref{fig:env}. 
A very good stability of the mean temperature value over several days has been observed with variations $<$1$^{\circ}$C, while slightly  bigger fluctuations have been evidenced on a shorter time scale ($\sim$h).  Such stability of the temperature, along with  the humidity and the pressure, should provide optimal conditions for the detector and the electronics equipment during our future data taking. 

A similar medium term stability can be evidenced for the radon concentration in the air with a total mean value of $\sim$80 Bq/m$^3$, however significant fluctuations (up to a factor 3) can be observed within few hours. While no special warning can be issued at the moment, we plan to continue our Rn monitoring in the future.

\section{Neutron background measurements}
\label{sec:7}

A complete characterisation of the neutron background is an important issue for dark matter experiments. The neutrons can produce single scattering nuclear recoils in the energy region of interest (ROI), thus becoming an irreducible background for these extremely sensitive experiments. Neutrons are produced through several mechanisms. Natural radioactive isotopes such as $^{238}\text{U}$ and $^{232}\text{Th}$, which are constituents of the laboratory rock walls and the detector materials, generate fast neutrons in the decay chains via spontaneous fission and through ($\alpha$,$n$) reactions. In addition, neutrons of very high energies are generated from cosmic-ray muon-induced spallation reactions. For this reason, the ArDM underground operation at LSC needs to be completed with site-specific neutron background measurements.

There have been measurements of the neutron background at LSC.  A first measurement was performed at the old LAB2500, as part of the IGEX-DM experiment. A value of  $\phi=3.82(44)\times10^{-6}\text{cm}^{-2}\text{s}^{-1}$ was estimated, due essentially to radioactivity-related neutrons coming from the rock \cite{Carmona}. A second measurement has been performed by the CUNA collaboration 
at Hall A with six large $^{3}\text{He}$ proportional tubes, each one embedded in a polyethylene block of different thickness. The flux distribution as a function of neutron energy was determined in the range from 1 eV to 10 MeV. The value of the flux integrated over this energy range is $\phi = (3.47\pm0.35)\times10^{-6}\text{cm}^{-2}\text{s}^{-1}$. 

In ArDM, we foresee specific neutron background measurements at LSC with a BC501A liquid
scintillation detector, similar to the one performed in reference \cite{BC501A,BackmeasurementsCPL},
and a $^{3}\text{He}$ proportional counter for monitoring purposes \cite{Jordan}. 
The data acquisition system, composed by standard NIM modules and a CAEN digitizer, and the previously described detectors are currently being tested at CIEMAT. Regarding the software, a data acquisition program for CAEN digitizer management and specific analysis software have been developed at CIEMAT. We are currently working in some improvements for the analysis routines in order to optimise the neutron/gamma separation.

\begin{figure}[htbp]
\begin{center}
\includegraphics[height=12.2pc]{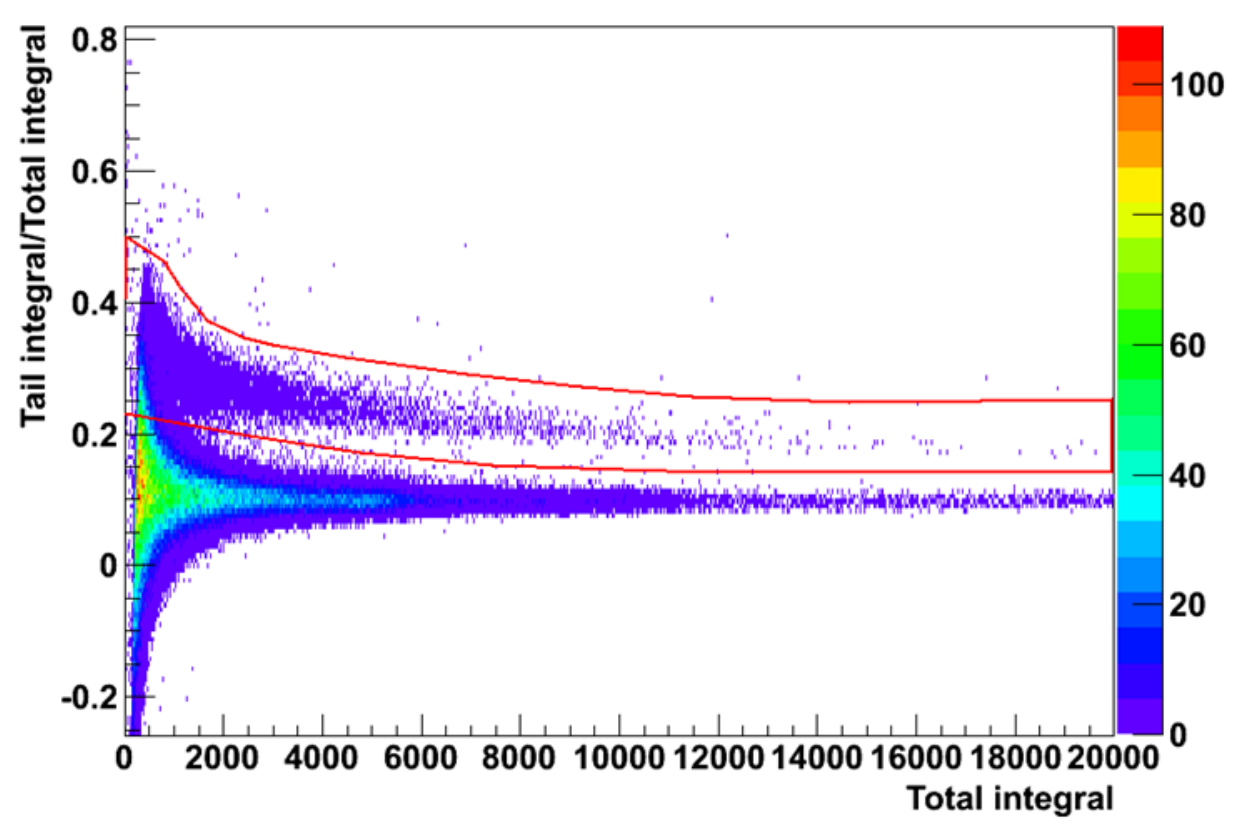}
\includegraphics[height=12pc]{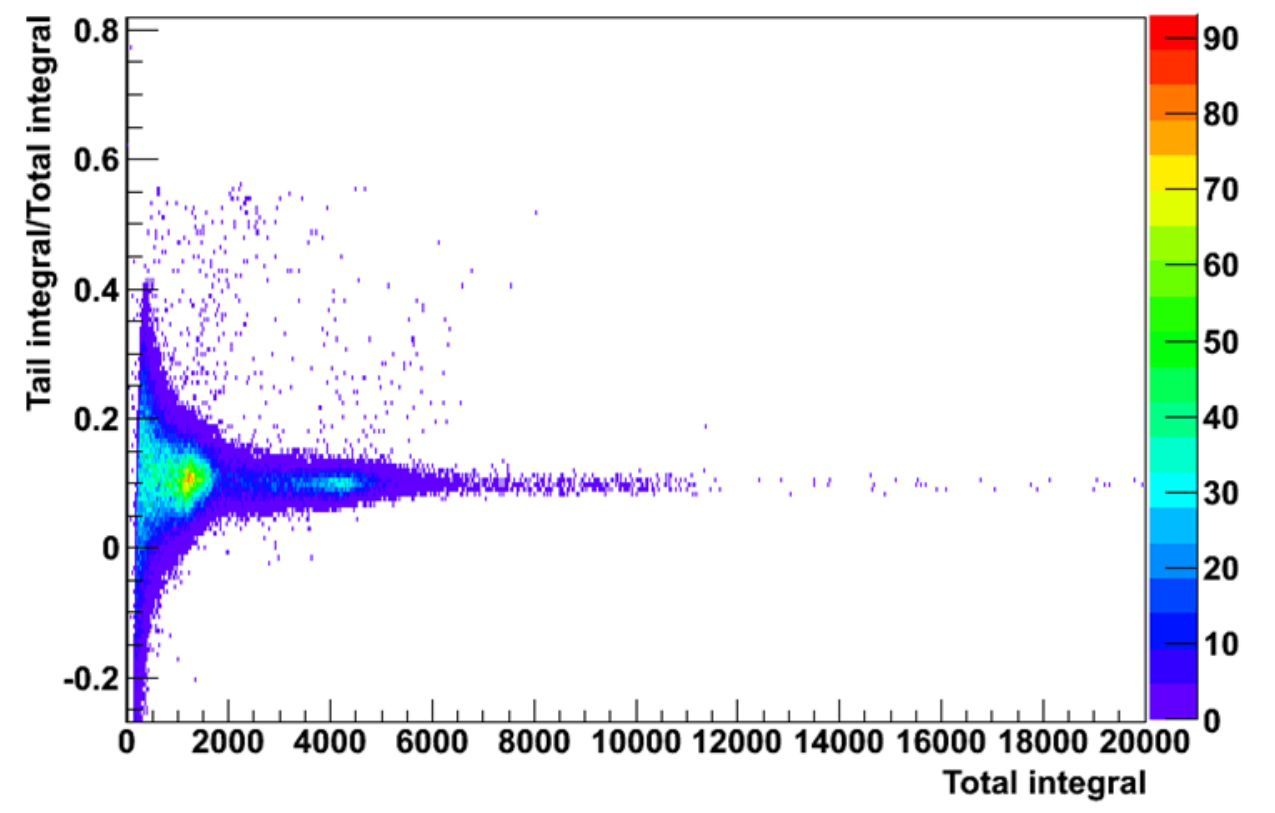}
\caption{(Right) ratio between tail integral and total integral vs total integral calculated from $^{252}\text{Cf}$ signals (neutron and gamma source), where  1 MeVee corresponds to approximately 4200 units of area. The region where the neutron events are located is surrounded by a red line. (Left) same with $^{22}\text{Na}$ (only gamma source).}
\label{Neutrongammaregions}
\end{center}
\end{figure}

First data taken with the apparatus and $^{252}\text{Cf}$ and $^{22}\text{Na}$ sources at CIEMAT
have been analysed. The results show that it is possible to distinguish between signals produced by neutrons and gammas by using the digital charge integration (DCI) discrimination method. This is due to the fact that,  neutron signals show a longer tail than the gamma ones. As a consequence, the ratio between the tail integral and the total integral is larger for neutrons than for gammas. If we compare the results obtained with  $^{252}\text{Cf}$ source (neutron and gamma source) with the ones obtained with $^{22}\text{Na}$ (gamma source), we clearly identify the signals due to gammas and confirm that we have correctly selected the neutron events (see \Cref{Neutrongammaregions}).

\begin{figure}[hbtp]
\begin{center}
\includegraphics[height=15pc]{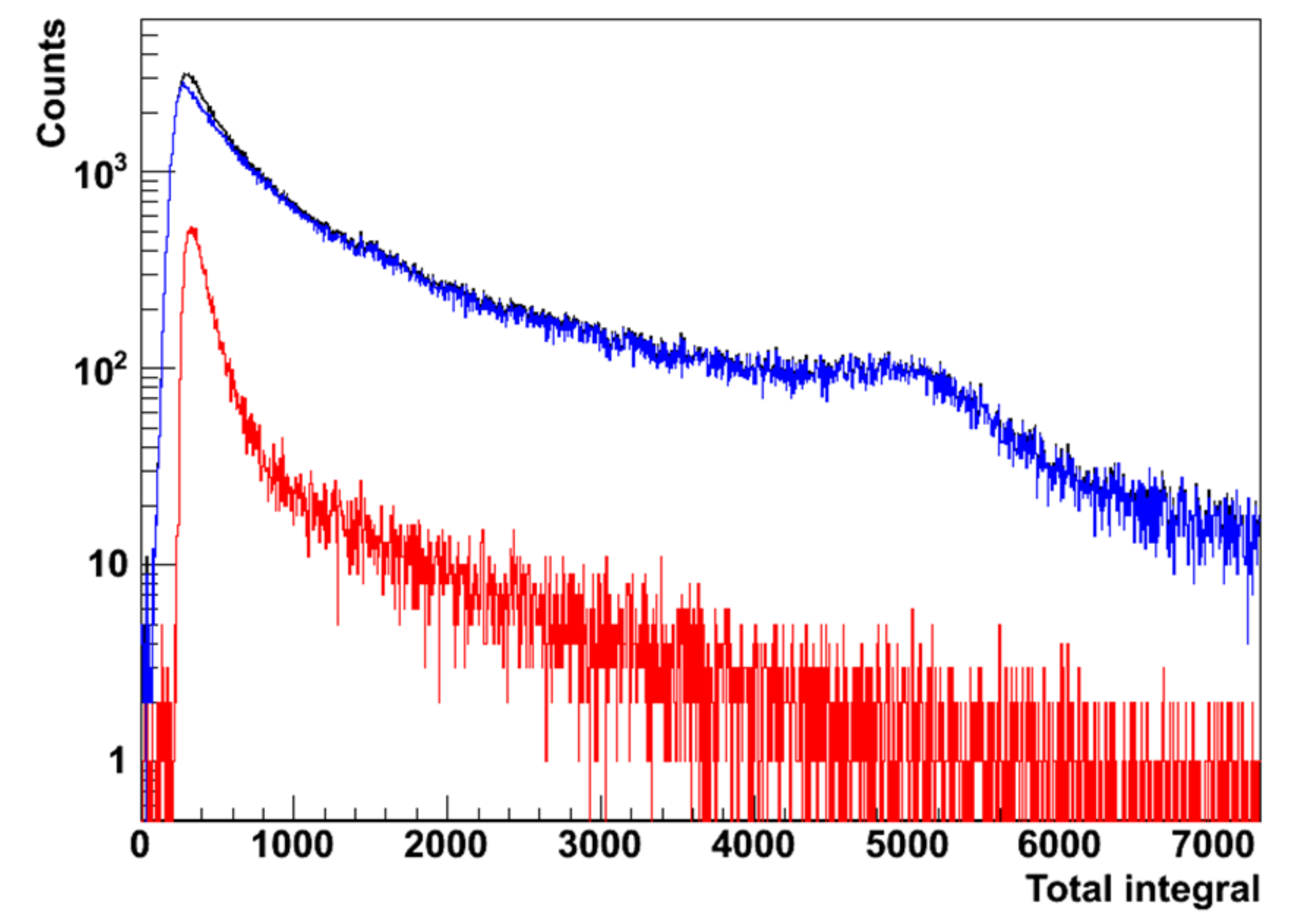}
\caption {Neutron spectrum from  $^{252}\text{Cf}$ source. The total spectrum is depicted in black, the blue spectrum corresponds essentially to the gamma spectrum and in red we have plotted the neutron spectrum (it may contain some gammas due to the superposition of the neutron and gamma regions at low energies).}
\label{NeutronspectrumCf252}
\end{center}
\end{figure}

If we select the region where the neutron events are located, it is possible to obtain the fast neutron spectrum for neutron energies higher than 1 MeV (\Cref{NeutronspectrumCf252}).

Different radioactive sources are required for calibration of the detector, standard gamma-ray calibration sources, such as $^{137}\text{Cs}$,  $^{22}\text{Na}$,  $^{88}\text{Y}$ and  $^{208}\text{Bi}$ can be used. For characterising the neutron/gamma separation,  a $^{252}\text{Cf}$ or an Am/Be source will be used.
Moreover, a periodic calibration run with gamma ray sources will be necessary for monitoring the gain stability, which can be affected by temperature variations; although, they should not be very significant at LSC.

\section{Next steps towards full operation of ArDM}
\label{sec:8}

After the first commissioning of the PMTs with argon gas as described in \Cref{sec:3}, the next step is to be able to have continuous runs without the necessity of changing the gas. 
This requires the use of the gas purification system which was already successfully
operated at CERN during the commissioning runs of ArDM. 
It now will be transported to LSC and installed according to \Cref{sec:8.1}.

In a first step of operation with LAr, the cooling system will be tested. The necessary work steps are described in \Cref{sec:8.2}. This operation will give us the possibility to measure with cold gas instead of gas at room temperature. Due to the higher density at equal pressure it will be possible to increase the energy resolution and to calibrate the detector with external gamma sources, instead of the internal alpha source.

\subsection{Installation of GAr recirculation system}
\label{sec:8.1}

\begin{figure}[hbtp]
\begin{center}
\begin{tabular}{cc}
\includegraphics[height=70mm]{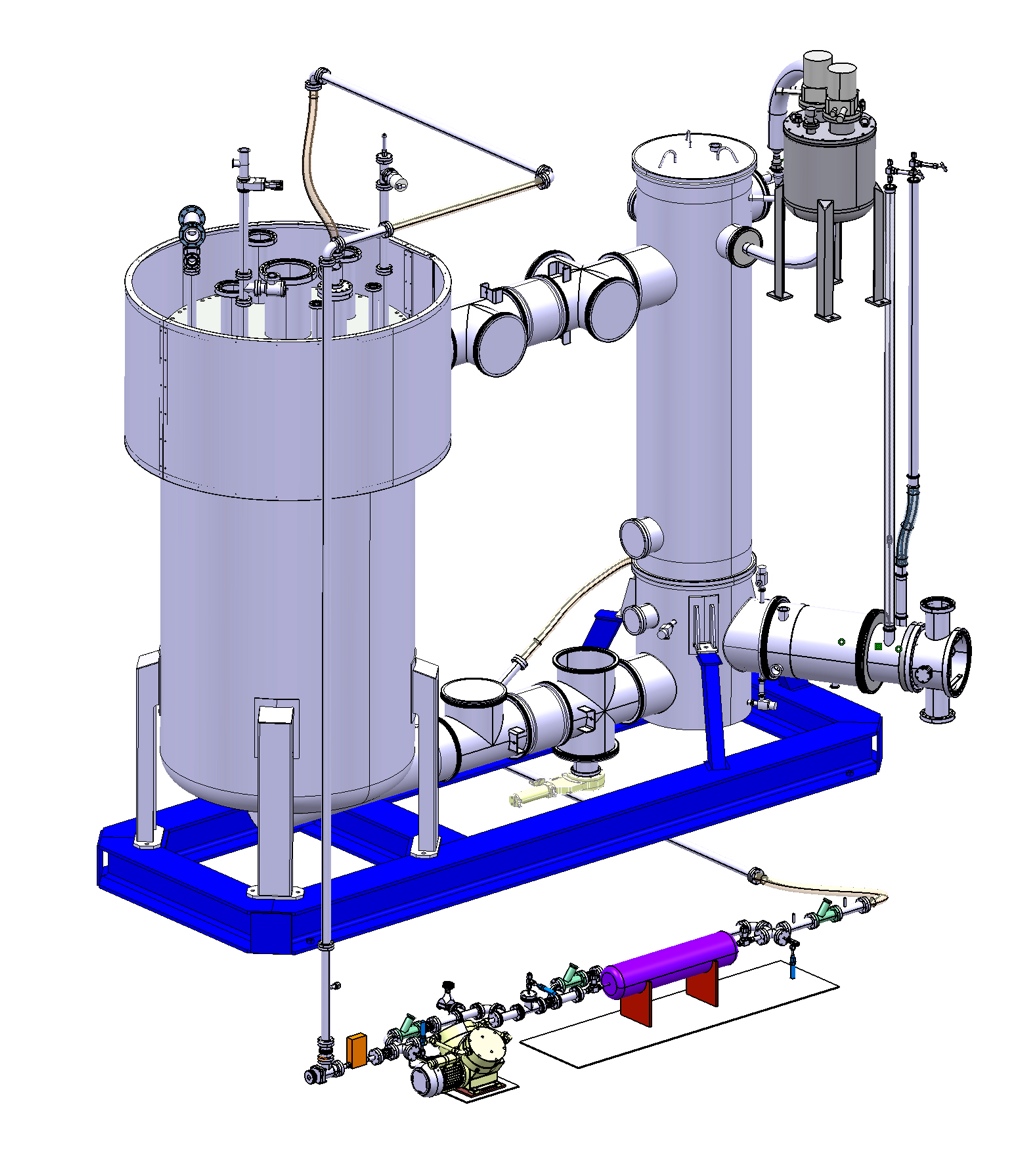}&
\includegraphics[width=95mm]{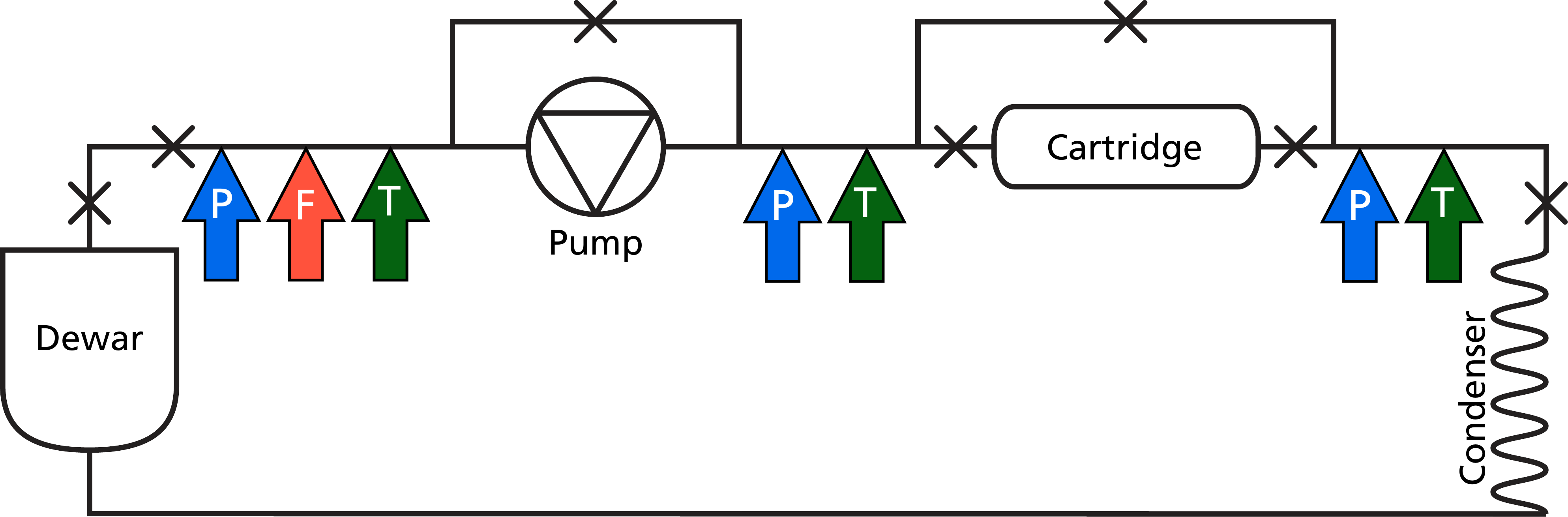}
\end{tabular}
\caption{Left: CAD model of the planned gas recirculation system. The gas is extracted from the main dewar from the top flange and pumped through a purification cartridge. Then, passing a condenser, it is brought back to the dewar from the bottom. Right: Schematic view of the circuit: The complete process is monitored by three pressure- (blue) and three temperature sensors (green), as well as a mass flow meter (red).}
\label{f_GasRecirculation}
\end{center}
\end{figure}

The gas argon recirculation system has been built and tested at CERN. 
\Cref{f_GasRecirculation} shows on the left the CAD drawing of the planned installation of it at LSC. 
The argon is extracted from the detector at the top flange (the gas vapour phase in case of operation with liquid). It then is pushed through a commercial SAES MicroTorr MC4500 gas filter by a double membrane pump with a maximum flow of about \unit[200]{slpm}.
The purified gas is brought back to the detector, entering the dewar from the bottom in order to get a maximum mixing of the gas. In case of operation with LAr, the purified gas is re-condensed. The necessary heat exchanger is part of the cooling system of the detector and uses the same cryocoolers. It is already installed at LSC.
The entire process is monitored and controlled by several pressure sensors and temperature probes. In the right part of \Cref{f_GasRecirculation}, a schematic illustration of the instrumentation is shown. The gas flow is monitored by a mass flow meter (MFM), situated in front of the pump. All sensors are connected to the PLC safety system and in case of irregularities, the pump is switched off automatically.

\subsection{LAr bath cooling test} 
\label{sec:8.2}

After the full commissioning of the gas recirculation system as described in the previous section 
we plan to start commissioning 
our system with a  small amount of about \unit[200]{L} of liquid in the cooling bath. 
This step is important to find possible cold leaks in the main vessel, as well as to get accustomed to the actual operation at cryogenic temperatures. We also plan to fill the detector with gas argon at \unit[87]{K}. 
The impact of the cryocoolers on the air-conditioning of the laboratory will be tested.

Once the vessel has reached a low temperature, 
we would like to fill it with $\unit[\sim1]{bar}$ of argon gas, 
cooled down to $\unit[\sim 90]{K}$ (density $\rho\approx 5.2~\unit{kg/m^{3}}$). 
The increased density, compared to gas at room temperature, gives a larger 
cross section for low energetic particles and measurements of external 
sources will be easier and more precise. 

During this test and also in the future, 
we plan to condense gaseous argon delivered in gas bottles. This is a slow process and, according to our experience, it takes about one week to fill the cooling jacket with LAr and to cool down all the steel and instrumentation.  \Cref{t_Filling_Bath} gives the most important numbers related to this operation.
\begin{table}[htdp]
\begin{center}
\begin{tabular}{cc}
\hline
Volume of the cooling jacket:&$\unit[\sim 200]{L}$\\
Condensation rate: & $\unit[1.3]{l/hr}$ of LAr\\
Total time: & $\unit[\sim 6.5]{days}$\\
Total amount of liquid per gas bottle:& \unit[12.5]{l/bottle} (\unit[150]{l/battery})\\
Total number of needed batteries: & 2 (replacement after $\unit[\sim5]{ days}$) \\
\hline
\end{tabular}
\end{center}
\caption{Numbers, related to filling the cooling bath of ArDM.}
\label{t_Filling_Bath}
\end{table}

We think that it is better to condense gaseous argon instead of bringing in LAr to the laboratory. Liquid always undergoes a pressure drop what causes evaporation of argon. Also the initial cooling of all the steel liberates a lot of energy and the cryocooler are not proportionated to handle this amount of gas in such a short time. The use of liquid therefore requires the constant release of argon gas, while condensation is possible in a completely closed and sealed system.

\section{R\&D for alternative methods of WLS coating and fixing}
\label{sec:9}

Currently the ArDM PMTs and the TTX reflectors are coated with TPB by vacuum evaporation deposition. 
While this coating method provides the best light yield for detecting LAr scintillation, the deposited TPB layer is typically relatively fragile and requires careful handling to avoid abrasion. In addition, effect of such a feature on the long-term stability of the WLS performance is a subject which still requires a clearer understanding. 

An R\&D program is underway to investigate 
alternative methods of WLS coating in view of  long-term stability of light yield. A first subject is to make the coating layer durable with good adherence to substrate and high resistance to mechanical abrasion, without weakening the WLS properties. 

Such a configuration with a WLS-coated transparent polymer window in front of a PMT was chosen as a baseline design for the new R\&D. 
The choice of new materials as optical interfaces has required special attention due to the thermodynamic conditions and low temperatures. 
The different substrates used as windows have been tested first under cryogenic conditions to verify their behaviour. 
Within the large family of polyesters transparent to near UV and compatible with the chemical and physical characteristics of the WLS, we selected a substrate suitable to deposit a homogeneous and stable layer of TPB. 
The research has thus been oriented to polycarbonates: 
Makrolon$\unit{^{\mbox{\tiny{\textregistered}}}}$ and Lexan$\unit{^{\mbox{\tiny{\textregistered}}}}$\footnote{Produced by Bayer AG and SABIC, respectively.} 
that offer a wide range of innovative solutions with and without protection for UV light. 
The substrates were cooled down to the liquid helium temperature and the morphological characteristics and optical properties of the deposits made with the different methods have been repeatedly measured.

\begin{figure}[htbp]
\begin{center}
\includegraphics[height=17pc]{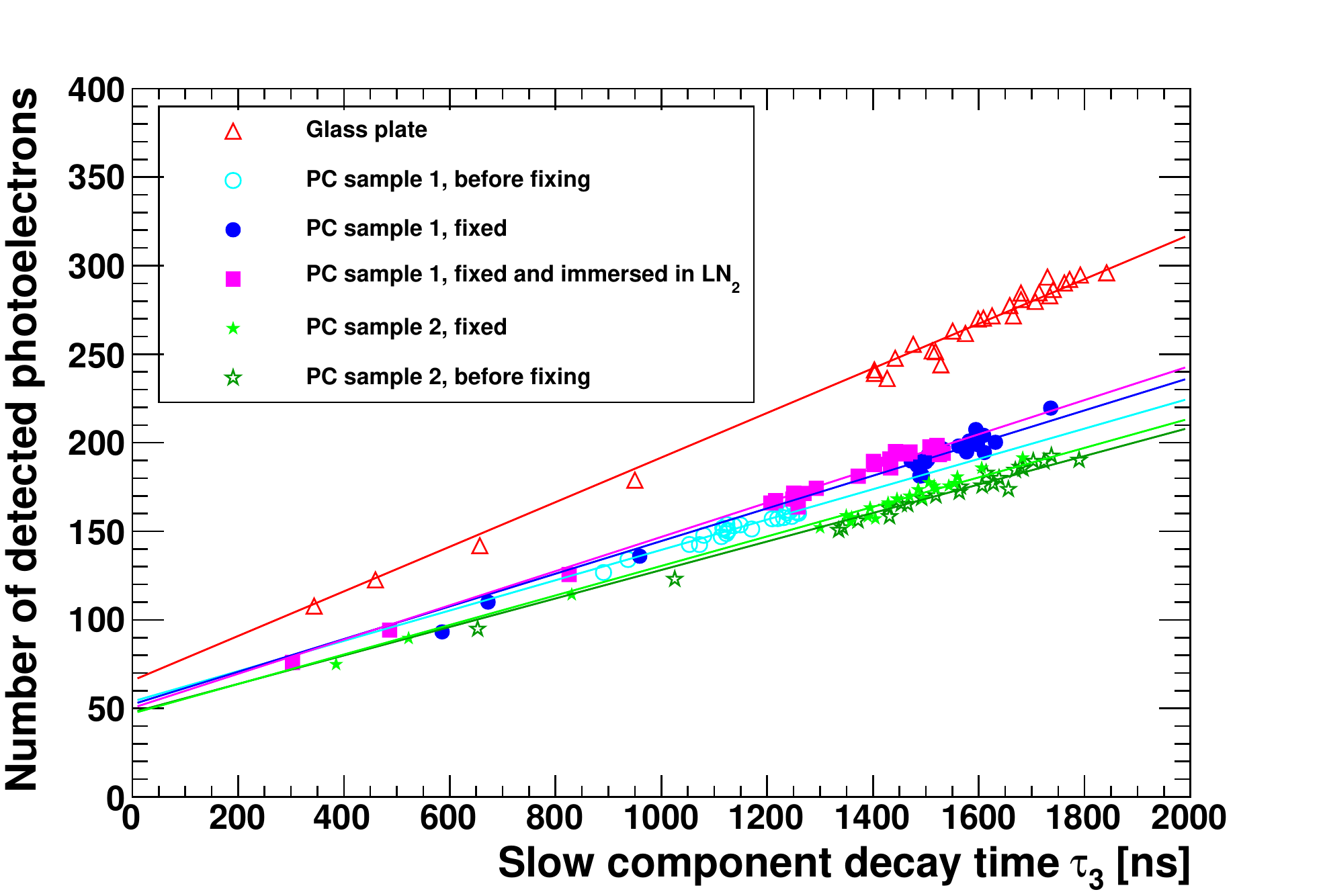}
\caption{Light yield measured with a window of different types, coated with TPB by evaporation. Gas argon scintillation light induced by alpha's from $^{241}$Am was measured by a 3'' high-QE PMT (Hamamatsu R11065) incorporated with one of the windows, and is plotted as a function of $\tau_3$, the decay time of the slow scintillation component (see \Cref{sec:3}). The polycarbonate samples were measured before and after chemical fixing, and also after 
having been immersed in liquid nitrogen (PC sample 1).}
\label{fig:makrolon}
\end{center}
\end{figure}

A preliminary result is presented in \Cref{fig:makrolon}. 
In a small gas argon test chamber, 
a 3'' high-QE PMT (Hamamatsu R11065) was installed with a 
3-mm-thick polycarbonate (PC) disc 3'' in diameter attached to its window. 
Argon scintillation light (128 nm) induced by alpha particles from an $^{241}$Am source hit the TPB layer and was converted to visible blue light, a fraction of which traversed through the sample window and finally was detected by the PMT. 
The PC disc was coated with TPB by evaporation, and then was chemically fixed to the substrate with good adherence and high resistance to mechanical abrasion. 
The light yield was measured before and after the chemical fixing, and the thus measured number of photoelectrons is plotted as a function of $\tau_3$, the decay time of the slow scintillation component (see \Cref{sec:3}). 
Two PC samples were used: sample 1 without UV protection, i.e. having a good transmittance down to $\sim$350 nm, and sample 2 with UV protection having a sharp cut off at around 400 nm. 
For both samples, the measured light yield did not change by the chemical fixing. 
After fixing the sample 1 was measured also after 
having been immersed in liquid nitrogen, showing no change in its optical properties. 
As a reference, a curve measured with a 1-mm-thick glass plate coated with TPB by evaporation also is shown in the same plot. Somewhat lower light yield obtained systematically for the PC samples can partially be explained by a higher trapping efficiency for the re-emitted light inside the window, due to a larger refractive index of PC than that of glass, i.e. $\sim$1.6 and $\sim$1.5, respectively. It may also be due to a slightly less transmittance of PC for visible and near UV light, which needs to be clarified.  

Besides, a program for investigating aging of the prepared samples has been started with a dedicated setup and within six months it will be possible to verify the evolution of the efficiencies of the various optical solutions.

\section{Conclusion} 
Since more than a year, the ArDM setup has been constantly in ``vacuum phase'' operation at LSC.
This has allowed us to acquire significant experience on the lab conditions.
A major milestone has been achieved 
with the installation of the re-designed detector
with two completely new PMT arrays and 24 PMTs in total, 
and a freshly coated reflector. Electronics and DAQ has been installed
and operated.

Preliminary data collected with gas argon have allowed the
assessment of the performance of the newly rebuilt light collection system.
We observe a significant improvement of the light yield.
We now see at least 2 p.e./keVee with zero electric field in LAr.
We observe an improved uniformity of the total light yield over the entire height of the fiducial volume, 
when the light detected by the top and the bottom arrays are summed up. 

The present detector allowed recording events with 20 p.e. from the argon 
scintillation, clearly separated from noise dominated by dark counts. 
With a light yield 
of 2 p.e./keVee in LAr at zero electric field, it corresponds to 10 keVee, which  
is already close to our proposed energy threshold for WIMP searches.

Continuous temperature and radon monitoring have been performed. An in-situ
campaign of neutron background measurement is being planned.
Next steps involve the installation of the gas recirculation system and the 
test of the cryocoolers. The risk analysis specific to operation in LSC
is being conducted.

An R\&D is underway to make the WLS coating layer durable with good adherence to substrate and high resistance to mechanical abrasion for long-term stability of the WLS performance. 
Preliminary results demonstrated that a thin TPB layer deposited on a transparent polycarbonate window by evaporation can be stabilised by a chemical treatment without 
changing its optical properties. 

\section*{Acknowledgements} 

We take the opportunity to thank the Director and the personnel of Laboratorio Subterr\'{a}neo de Canfranc (LSC)
for their support.
We thank the Thin Film \& Glass group of CERN for the TPB coating of our PMTs. 
We acknowledge the kind support of the NEXT Collaboration by making available the evaporator for coating of the side reflectors. 
We are grateful to The Nuclear Innovation Unit at CIEMAT and the rest of the CUNA Collaboration for their participation and support in the neutron background measurements campaign. 
We thank also the department ``Lunghezza and Fotometria'' of the institute INRIM, Torino, for all their support. 



\begin{thebibliography}{99}

\bibitem{Steigman:1984ac}
  G.~Steigman and M.~S.~Turner,
  ``Cosmological Constraints on the Properties of Weakly Interacting Massive Particles,''
  Nucl.\ Phys.\ B {\bf 253} (1985) 375.
   
\bibitem{Rubbia:2005ge}
  A.~Rubbia,
  ``ArDM: A Ton-scale liquid Argon experiment for direct detection of dark matter in the universe,''
  J.\ Phys.\ Conf.\ Ser.\  {\bf 39} (2006) 129
  [hep-ph/0510320].
  
\bibitem{Marchionni:2010fi} 
  A.~Marchionni {\it et al.}  [ArDM Collaboration],
  ``ArDM: a ton-scale LAr detector for direct Dark Matter searches,''
  J.\ Phys.\ Conf.\ Ser.\  {\bf 308}, 012006 (2011)
  [arXiv:1012.5967 [physics.ins-det]].
  
\bibitem{Epprecht:2012Diss}
  L.~Epprecht, 
  ``Design, construction and first commissioning of the ArDM detector,''
  Diss., ETH Z\"{u}rich, Nr. 20921 (2012).
  %

\bibitem{Boccone:2009kk}
  V.~Boccone {\it et al.}  [ArDM Collaboration],
  JINST {\bf 4} (2009) P06001
  [arXiv:0904.0246 [physics.ins-det]].

\bibitem{Degunda:2013Diss}
  U.~Degunda, 
  ``Measurement of the electronic recoil contamination and development of the control system for the ArDM experiment,''
  Diss., ETH Z\"{u}rich, Nr. 20966 (2013).


\bibitem{Lazzaro:2012Diss}
  C.~Lazzaro, 
  ``Reconstruction of the muon tracks in the OPERA experiment and first results on the light collection in the ArDM experiment,''
  Diss., ETH Z\"{u}rich, Nr. 20551 (2012).

\bibitem{Carmona}
J.M. Carmona et al., Astrop. Phys 21 (2004) 523.

\bibitem{BC501A}
C. Guerrero, D. Cano-Ott, M. Fern\'andez-Ord\'o\~{n}ez, E. Gonz\'alez-Romero, T. Mart\'inez, D. Villamar\'in, \emph{Analysis of the BC501A neutron detector signals using the true pulse shape},
Nuclear Instruments and Methods in Physics Research NIM A597 (2008) 212--218.

\bibitem{BackmeasurementsCPL}
H.J. Kim et al., \emph{Measurement of the neutron flux in the CPL underground laboratory and simulation studies of neutron shielding for WIMP searches}, Astrop. Phys. 20(5):9 (2004).

\bibitem{Jordan}
D. Jordan, J.L. Ta\`in et al., \emph{Measurement of the neutron background at the Canfranc Underground Laboratory LSC}, Astroparticle Physics 42 (2013) 1--6.


\end{thebibliography}
\end{document}